\documentclass[preprint,showpacs,amsmath,amssymb,floatfix,prb,aps]{revtex4-1}
\usepackage{graphicx}
\usepackage{dcolumn}%
\usepackage[english]{babel}
\usepackage{amsmath}
\usepackage{amssymb,amsfonts,textcomp}
\usepackage{supertabular}
\usepackage{bm}
\usepackage{setspace}
\newcommand{\bigO}{\ensuremath{\mathcal{O}}}

\begin{document}
\title{Kinetics of liquid-solid phase transition in large nickel clusters}
\author{Alexander V. Yakubovich, Gennady Sushko, Stefan Schramm, and Andrey V. Solov'yov}

\affiliation{
Frankfurt Institute for Advanced Studies, Ruth-Moufang-Str. 1,
60438 Frankfurt am Main, Germany\\
}
\date{\today}
\begin{abstract}
\noindent 

In this paper we have explored computationally the solidification process of large nickel 
clusters. This process has the characteristic features of the first order phase transition 
occurring in a finite system. The focus of our research is placed on the elucidation of
correlated dynamics of a large
ensemble of particles in the course of the nanoscale liquid-solid phase transition
through the computation and analysis of the results of molecular dynamics (MD) simulations
with the corresponding theoretical model. This problem is of significant interest and importance, because the controlled dynamics of systems 
on the nanoscale is one of the central topics in the development of modern nanotechnologies.

MD simulations in large molecular systems are rather computer power demanding.
Therefore, in order to advance with MD simulations we have used modern 
computational methods based on the graphics processing units (GPU). The advantages of the use of GPUs
for MD simulations in comparison with the CPUs are
demonstrated and benchmarked. The reported speedup reaches factors greater than 400. 
This work opens a path towards exploration with the use of MD of a larger
number of scientific problems inaccessible earlier with the CPU based computational technology.

\end{abstract}

\maketitle

\section{Introduction}

The melting process in a macroscopically large system occurs at a certain temperature
under fixed external pressure. This process is a first order phase transition and it
manifests itself as a spike in the heat capacity of the system at the transition
temperature. The reverse process of freezing or solidification occurs
at a temperature below the thermodynamic melting point due to nucleation
nature of the solidification phase transitions~\cite{Aguado11}.

In small systems the finite volume that is available for the precursor
formation alters the kinetics of the solidification phase transition.
 The study of supercooled metal droplets of the size
of several micrometers was performed in early 1950s~\cite{Turnbull50, Turnbull52}. In Ref.~\cite{Turnbull52} it was shown
that the mercury droplets of 2 -- 8 $\mu$m in diameter solidify at rates that are proportional to the droplet volume.
In the systems of the {\it nano}meters size the ratio of the surface to volume
atoms increases and the reduced binding energy of the surface atoms changes 
significantly the total energy of the system.
The first thermodynamic model bridging the melting point of the clusters with their
size was proposed by Pawlow more than a century ago~\cite{Pawlow1909}. 
In multiple subsequent works it
was confirmed that the melting temperature of a small spherical particle decreases
with the reduction of its radius~\cite{Buffat76,Castro90,Lai96,Bottani90}. 
In small particles the relative fraction of the 
surface atoms is higher, which leads to the decrease of the melting temperature.
This size effect  have been confirmed for the clusters having diameters down to 2~nm~\cite{Buffat76}.
Its explanation is based on the
characteristic radial dependence of the ratio of the surface to volume energy of a finite system.

However, for clusters having sizes smaller than 1 - 2 nm,
the melting temperature is no longer a monotonic function of
the cluster size. Experiments on sodium clusters Na$_N$ with
number of atoms N=50 -- 360 have demonstrated that the melting
temperature as a function of size shows a prominent irregular
structure with local maxima~\cite{Haberland97,Haberland98,Haberland99,Haberland05}. The origin of
the nonmonotonic variation in the melting temperature with
respect to cluster size lies in the interplay between electronic
and geometric shell effects in the sodium clusters
and the entropy change in the course of such nanoscale phase transitions~\cite{Haberland05}.

Intensive
theoretical efforts have been undertaken to identify the details
of the geometric and electronic structures underlying the
variations in the melting temperature~\cite{Calvo00,ReyesNava03,Calvo04,Manninen04,Chacko05,Aguado05a,Aguado05b,Noya07}.
Experiments on small ion clusters of tin~\cite{Shvartsburg00} and gallium~\cite{Breaux03} 
have confirmed the violation of the linear relationship between
the reduction in the melting temperature and the inverse
radius of the cluster. It was discovered that the melting
temperature of selected Sn$_N$ and Ga$_N$ clusters, of sizes N $<$ 40,~can
considerably exceed the melting temperature of the
corresponding bulk material~\cite{Shvartsburg00,Breaux03}. This behavior was explained
by the structural differences between the small clusters
and the bulk~\cite{Shvartsburg00}.

There are also other factors which affect the melting temperatures of small clusters.
For instance, impurities play a significant role. Thus, in Ref.~\cite{Lyalin09} it was demonstrated that
a single atom impurity leads to significant changes in the thermodynamic properties of the Ni$_{147}$ cluster.
Also, the melting temperatures differ significantly in nanoalloys as compared to pure materials.
In Ref.~\cite{Ding04} it was shown that alloying iron clusters consisting of up to 2400 atoms with carbon reduces their melting temperature by 100 -– 150 K at a carbon concentration of 10$\%$ –- 12$\%$. Recently, the thermal behaviour of free and alumina-supported Fe-C nanoparticles has been investigated~\cite{Cutarolo07}. It was observed that the presence of the substrate raises the melting temperature of medium and large Fe$_{(1 −- x)N}$C$_{xN}$ nanoparticles (x=0 -- 0.16, N=80 $–-$ 1000) by 40 -- 60 K~\cite{Cutarolo07}.
In Ref.~\cite{Petrov:2006gf} the freezing-melting hysteresis associated with a free energy barrier between solid and liquid phases was investigated theoretically for the materials in pores.

There are many papers devoted to the computer simulations of the melting process of metal clusters consisting of up to few hundreds of atoms.
For the small systems one can utilize {\it ab initio} approaches based on Density Functional Theory~\cite{Ruban:2008ko} or more
empirical Car-Parinello~\cite{Aguado:2001ie} or Tight-Binding schemes~\cite{Goringe:1999bx}. The aforementioned methods are capable of reproducing relatively
accurately the structural and energetic properties of the clusters. However, computational complexity of {\it ab initio}
methods does not allow for modelling of metal clusters consisting of thousands of atoms on the time scales exceeding nanoseconds.  

Investigations of melting of clusters of several thousands of atoms are carried out using empirical classical
potentials for the description of interatomic interactions. The delocalized nature of $d$-electrons in the transition metal clusters implies utilization of so-called many-body potentials, which account for two- and three-body interatomic interactions.
The most widely used potentials for the description of interactions between nickels atoms are Finnis-Sinclair and Sutton-Chen potentials. In Ref.~\cite{Qi:2001jt} MD simulations of melting and solidification of Ni nanoclusters with up to 8007 atoms are performed with the use of the classical quantum--corrected Sutton-–Chen potential~\cite{PhysRevB.59.3527}, and the characteristics of the liquid-solid phase transition are analysed for different cluster sizes. The investigation of melting and solidification
of Ni clusters consisting of up to 32000 atoms using the conventional Sutton-Chen potential is presented in~\cite{Shibuta:2010eh}.
In that paper the reduction of the melting and solidification temperatures for the face-centred cubic metals is reported to be
negatively correlated with the particle radius, and the Gibbs-Thomson coefficient is found to be proportional to the melting point.    
In Ref.~\cite{Shibuta:2011} the investigation of the effect of the cooling rate on the final structure of Mo$_{6750}$ clusters is performed with the use of the Finnis-Sinclair potential.

In the present paper we conduct MD simulations of Ni$_{2057}$ clusters on the timescale up to 65 ns and discuss the kinetics
of the solidification phase transition for the clusters as a function of the overcooling rate. 
From the analysis of MD simulations we extrapolate the thermodynamic properties for the clusters of
arbitrary size up to the bulk.  

Focus of our research is placed on the elucidation of
correlated dynamics of a large ensemble of particles in the course of the nanoscale liquid-solid 
phase transition. Obviously, this problem is of significant interest and importance, 
because the controlled dynamics of systems 
on the nanoscale is one of the central topics in the development of modern nanotechnologies.
For the purposes of this analysis we have chosen 
the melting process of large nickel clusters and
performed  a systematic theoretical
analysis of their dynamics in the course of melting and solidification.

The choice of Ni clusters for these studies is motivated by their high
chemical and catalytic reactivity, unique properties, and multiple
applications in nanostructured materials~\cite{Goddard07}. An important
example of such an application is the process of the catalytically
activated growth of carbon nanotubes.
The thermodynamic state of the catalytic nanoparticle plays a crucial role in the carbon
nanotube growth~\cite{Harutyunyan05}. The important question is whether
the catalytic nanoparticle is molten or frozen during the
nanotube growth process. It was demonstrated that when carbon
nanotubes are grown on large 3-4 nm iron nanoparticles
at temperatures lower than 1200 K, the catalytic particle is not completely molten~\cite{Ding04}. Thus, the mechanism of
nanotube growth can be governed by the surface melting of
the cluster. 
Therefore, the advanced MD simulations
of melting of large Ni clusters are important for a
reliable evaluation of the conditions at which the carbon nanotube growth process takes place~\cite{Lyalin09, Ding04, Obolensky07, Harutyunyan08}.
In order to achieve this task it is necessary to perform MD
simulations for relatively large nanoparticle sizes and large simulation times,
which imposes the use of the most advanced computational techniques based on the GPUs.
The advantages of this
technology for MD simulations in comparison with the CPU based one 
is demonstrated and benchmarked. The reported speedup reaches factors greater than 400. 
This work opens a path towards exploration of a larger number of scientific problems inaccessible earlier with the CPU based computational technology.

The CPU calculations of the nickel clusters were carried out using a 
multi-purpose computer code MesoBioNano Explorer (MBN Explorer)~\cite{MBNExplorer}.
MBN Explorer allows to use a broad variety of interatomic potentials, to model different molecular systems, such as atomic clusters, fullerenes, nanotubes, polypeptides, proteins, DNA, composite systems, nanofractals, etc. Despite the universality, the computational efficiency of MBN Explorer is comparable (and in some cases even higher) than the computational efficiency of other software packages.

For the purposes of this work we have adopted some of the MD algorithms of MBN Explorer to run them on GPUs.
In particular, we have rewritten the part of the code responsible for calculations of the Sutton--Chen potential and force using Open Computing Language (OpenCL). OpenCL is a framework for writing programs that execute across heterogeneous platforms consisting of CPUs, GPUs, and other processors. The details of the implementation are discussed in the following section. 

The paper is organized as follows. In section~\ref{sec:T&C} we introduce the Sutton--Chen potential, describe the theoretical model of nucleation and growth of solid state precursors in the course of the solidification phase transition, and present the details of the computational approach utilized in the work. In section~\ref{sec:R&D} we demonstrate the results
of MD simulations of Ni$_{2047}$ clusters and analyse radial distribution function, diffusion coefficients for
molten and solid states of the cluster, and the correspondence of MD simulation results with the theoretical model for the solidification rate. In section~\ref{sec:con} we draw conclusions to the paper.

\section{Theoretical and computational methods}
\label{sec:T&C}

\subsection{Interaction potential for nickel atoms}

The study of structural and dynamical properties of
transition-metal clusters is a challenging task due to the presence
of unfilled valence $d$ orbitals. The high density of the $d$
states and their delocalized character make the direct {\it ab initio}
methods computationally very demanding for clusters larger
than several tens of atoms~\cite{Nayak97}. In order to describe the structure
of clusters of larger sizes, one needs to use approximate
methods and model interatomic potentials.

An effective approach for the study of transition-metal clusters
is the embedded-atom method~\cite{Daw83,Daw84,Finnis84,Foiles86,Sutton90,Sutton96}, 
which takes into account
many-body effects. The latter appears through the inhomogeneous
electron density of the system. In this paper, an
MD study of nickel clusters has been
performed using the Sutton-Chen~\cite{Sutton90} many-body potential,
which belongs to the family of the embedded-atom types of
potentials. The Sutton-Chen potential~\cite{Sutton90} has been shown to
reproduce bulk and surface properties of transition metals
and their alloys with sufficient accuracy see, e.g. Refs.~\cite{Sutton91,Todd93,Lynden95,Nayak97,Doye98}
and references therein. The applicability of the
Sutton-Chen potential~\cite{Sutton90} to Ni clusters has been proven by the
direct comparison of the optimized structures and the binding
energies of small Ni clusters obtained within {\it ab initio}
method and with the use of the Sutton-Chen potential~\cite{Nayak97}.

The potential energy of the finite system within
the Sutton-Chen model has the following form:

\begin{equation}
\label{SC}
U_{pot}= \varepsilon \sum_i \left[ 
\frac{1}{2} \sum_{j \ne i} \left( \frac{a}{r_{ij}} \right)^{n} - c \rho_{i}^{1/2}  
\right] ,
\end{equation}

\noindent where

\begin{equation}
\label{eq:rho}
\rho_{i} =  \sum_{j \ne i} \left( \frac{a}{r_{ij}} \right)^{m}.
\end{equation}

\noindent Here $r_{ij}$ is the distance between atoms $i$ and $j$, 
$\varepsilon$ is a parameter with dimension of energy,
$a$ is the lattice constant, $c$ is a dimensionless parameter, and $n$ and $m$ are positive
integers with $n > m$. The  parameters provided by Sutton and Chen
for nickel have the following values:~\cite{Sutton90}
$\varepsilon = 1.5707 \cdot 10^{-2}$ eV,
$a = 3.52$ \AA\,
$c = 39.432$,
$n=9$, and $m=6$. 

\subsection{Theoretical model of the solidification kinetics}

In the liquid-solid phase transition the process of the formation of the new crystal phase is initiated
by the formation of the stable precursor of the solid state.
The energy of the formation of the precursor $E_p(r)$ of radius $r$ can be written as follows~\cite{LL5}:

\begin{eqnarray}
\label{eq:Prec}
E_p(r) = \frac{4 \pi r^2}{3}\left(3 \alpha -r \Delta f \right), 
\end{eqnarray}

\noindent where $\alpha$ is the liquid-solid surface tension coefficient and $\Delta f$ is the difference of the free energy densities for the liquid and the solid states. Here we consider the process of crystallization
from the overcooled liquid phase, therefore $\Delta f$ is positive and can be evaluated as follows:
\begin{eqnarray}
\label{eq:ddG1}
\Delta f &=& f_l - f_s,
\end{eqnarray}
\noindent where $f_l$ and $f_s$ are the free energy densities for the liquid and solid phases, correspondingly. $f_l$ and $f_s$ can be written as:
\begin{eqnarray}
\label{eq:ddG2.1}
f_l &=& e_{l0} +\frac{3NkT}{V_{cl}} - Ts_l, \\
\label{eq:ddG2.2}
f_s &=& e_{s0} +\frac{3NkT}{V_{cl}} - Ts_s, 
\end{eqnarray}
\noindent where $e_{l0}$, $e_{s0}$, $s_l$, $s_s$, $N$ and $V_{cl}$ are the ground-state energy densities of the liquid and solid phases,
entropies densities of the liquid and solid states, number of atoms in the cluster and cluster volume, correspondingly. The factor $3kT$ accounts for the energy stored in the thermal vibrations of the atoms at finite temperature. Using Eqs.~(\ref{eq:ddG2.1}-\ref{eq:ddG2.2}), Eq.~(\ref{eq:ddG1}) can be written as follows:
\begin{eqnarray}
\label{eq:ddG3}
\Delta f &=& \Delta s (T-T_0),
\end{eqnarray}
\noindent where $\Delta s=s_s - s_l$ and $T_0$ is the phase transition temperature. For the derivation of Eq.~(\ref{eq:ddG3}) we have accounted for the fact that free energy densities for both phases are equal at the phase transition temperature, i.e. $e_{s0} - T_0 s_s=e_{l0} - T_0 s_l$.

Eq.~(\ref{eq:Prec}) has a maximum at certain value of $r$, corresponding to the critical size $r_c$ of the precursor of the solid phase. If the
size of the precursor is larger than $r_c$ its further growth will reduce the free energy of the system, and the crystal phase will expand.
The value of $r_c$ corresponds to the size at which the derivative over $r$ of the Eq.~(\ref{eq:Prec}) is equal to zero, i.e.
\begin{eqnarray}
\label{eq:prec_crit}
r_c &=& \frac{2 \alpha }{\Delta f }.
\end{eqnarray}
\noindent Substituting $r_c$ from Eq.~(\ref{eq:prec_crit}) into Eq.~(\ref{eq:Prec}) one obtains the following value of the energy associated with the formation of the precursor of critical size:
\begin{eqnarray}
\label{eq:prec_crit_E}
E_p(r_c)=\frac{16 \pi \alpha ^3}{3 \Delta s^2 \left(T-T_0\right)^2},
\end{eqnarray}
where we have used Eq.~(\ref{eq:ddG3}) to substitute the expression for $\Delta f$.


Note that the precursors of the crystal phase are very improbable to be formed in the vicinity of the cluster surface, since
the surface atoms are in a liquid-like state even at the temperatures when the core of the cluster is in the solid state. 

The rate of the formation of the precursors of the solid phase can be calculated as follows:
\begin{eqnarray}
\label{eq:rate}
k = A \exp \left[-\frac{16 \pi \alpha ^3}{3 k_B T\Delta s^2 \left(T-T_0\right)^2}\right],
\end{eqnarray}
\noindent where $k_B$ is the Boltzmann constant and $A$ is the precursor formation frequency. 

\subsection*{Computational approach}
\begin{figure}
\includegraphics[width=1\textwidth,clip]{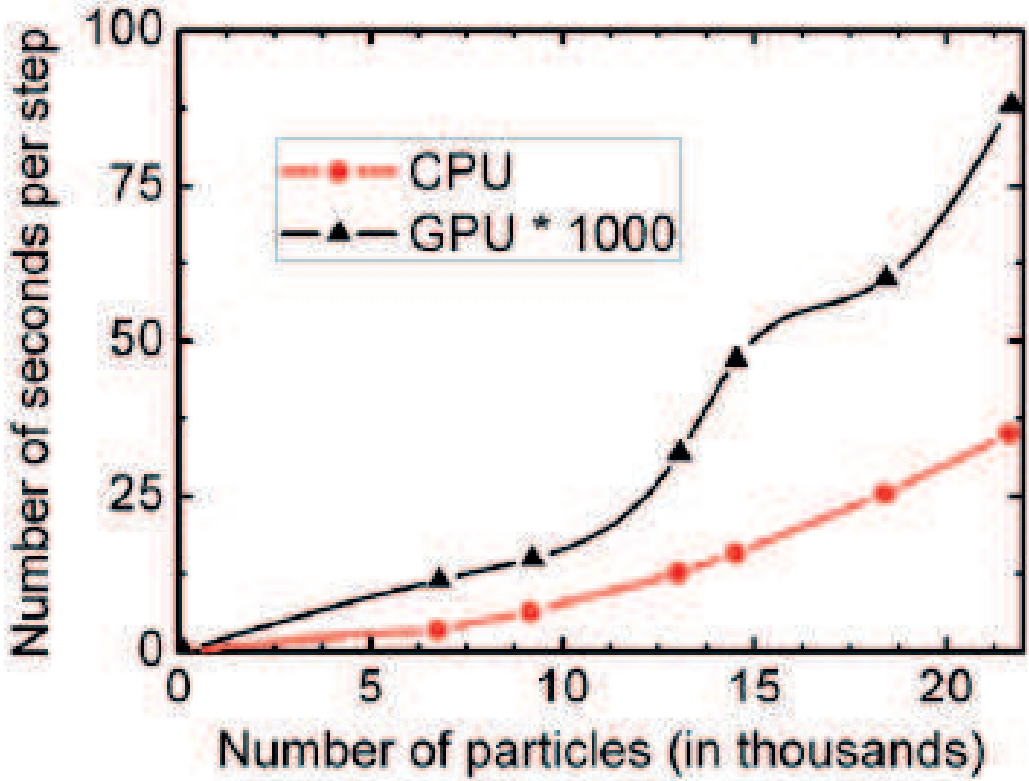}
\caption{ Dependence of computation time on the system size. From Fig. it is seen that the time is proportional to the squared number of particles, i.e. $\bigO (N^2)$. Note that the time per step is multiplied by a thousand for the GPU case.}
\label{fg:CT}
\end{figure}

The numerical solution of the MD equations becomes a challenging task
if one studies systems with a large number of atoms. Generally, the crucial time-consuming part
of the computation is the calculation of the forces between the particles at each time step.
For two-body forces this implies a computational effort rising with $N^2$, where $N$ is the number of atoms, see Fig.~\ref{fg:CT}.
An implicit many-body structure of the force as it is generated by the second term in the
Sutton-Chen potential, Eq.~(\ref{SC}), can potentially increase the calculation effort.
\\

As the force calculation essentially  implies the same type of calculation repeated many 
times over, this task is ideally suited 
to be implemented using a parallel-programming approach. If there are only few different 
types of forces involved, as is the case discussed here, a single-instruction multiple-data (SIMD) 
hardware environment can be adopted in a natural way, since the same instructions 
are used for all pairs of particles. Graphics cards (GPUs) fulfill these requirements 
in an ideal way. We therefore ported our original CPU based code 
MBN Explorer~\cite{MBNExplorer}
for CUDA as well as OpenCL programming frameworks. 
\\

We reduced the many-body part of the calculation to two successive two-body problems, in the first sweep calculating all $\rho_i$ 
in Eq.~(\ref{eq:rho}) and then determining the forces on the atoms.
\\

The calculations presented here were performed  on the LOEWE compute cluster 
at Frankfurt University, which features 778 graphics cards Radeon HD 5870.  
Each card includes 1600 streaming processors clocked at 850 MHz. 
The results GPU calculations were compared to a single core of Intel Core i7 870 CPU.

\begin{figure}
\includegraphics[width=1\textwidth,clip]{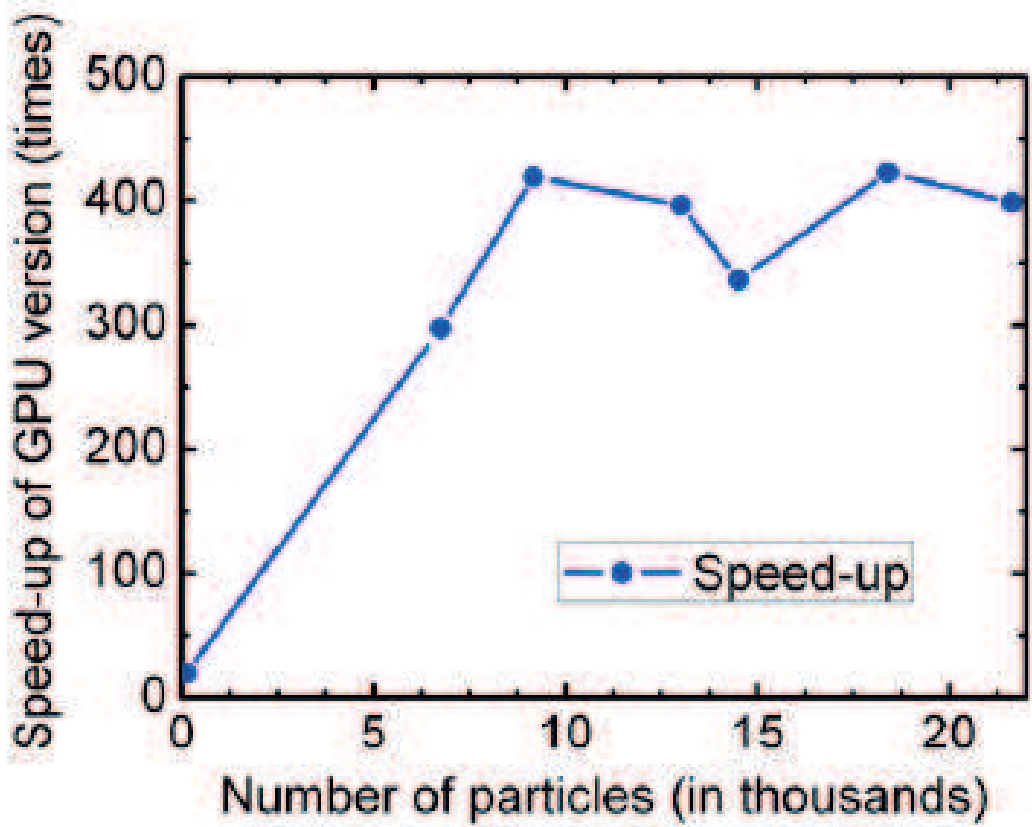}
\caption{ Speed-up of the computational time for GPU as compared to the CPU version.}
\label{fg:CT2}
\end{figure}

Fig.~\ref{fg:CT2} shows the speed-up of a MD simulation of nickel clusters of different sizes comparing the CPU and GPU versions of the code. As can be seen one can achieve speed-ups of more than two orders of magnitude for larger clusters. These results are consistent with theoretical estimation of performance of GPU and CPU in single precision. AMD Radeon specified to have 2720 GFLOPS in single precision and one core of Core i7 CPU can be estimated as nearly 15 GFLOPS in double precision. In double precision the difference would be less as AMD Radeon specified to have 544 GFLOPS in double precession. It means that for double precision code one can expect speed-up of nearly one hundred over single threaded code.

In
our code each particle was assigned to a certain thread. It implies that
if the number of particles is smaller then the number of stream processors GPU
will not be completely loaded. In order to utilize the GPU
efficiently the number of particles should be an integer
multiple of the number of the stream processors. 
This can be seen on speed-up plots presented in Figs. \ref{fg:CT}, \ref{fg:CT2}. Speed-up growth is nearly
linear until 10 thousands of particles. For the number of particles
exceeding 10 thousands the speedup saturates to a
constant level, corresponding to full utilization of the GPU. In the presented benchmarks the GPU
computation time was averaged over one thousand iterations in order avoid
accounting for random slowdowns and lags. CPU computation time was averaged over 10 iterations
since random slowdowns of CPU are expected to have minor influence on the computational time.


                              
Current calculations were performed as full N-body calculation without spatial cut-off for the interaction of the atoms, i.e., taking into account interactions between all atoms of the cluster. Therefore, in this work we have studied the cluster consisting of $\sim$2000 atoms, where a cut-off does not have a substantial advantage in computing time, but the
speedup due to utilization of GPUs is about 100.  The efficient implementation of interaction cutoffs in the GPU code is in progress and will be adopted and discussed in forthcoming work.

The MD simulations were carried out using a Verlet integrator with a time-step of 1~fs. The temperature control was carried out using a Langevin thermostat with a damping constant of 10~ps$^{-1}$. 

\section{Results and discussion}
\label{sec:R&D}


\subsection*{MD simulations of Ni$_{2047}$ clusters}

\begin{figure}
\includegraphics[width=1\textwidth,clip]{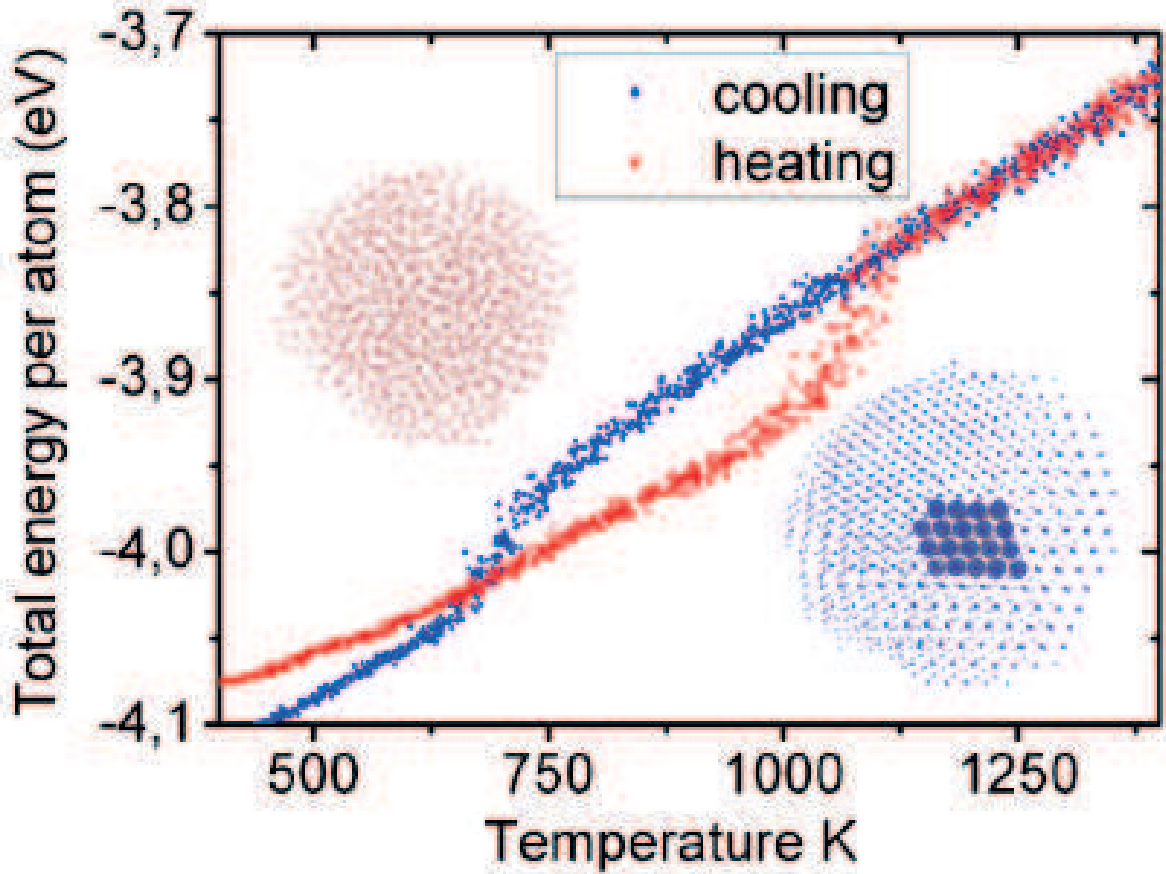}
\caption{Dependence of the
total energy per atom in heating (red) and cooling (blue) simulations. In the insets are shown the structures
of Ni$_{2047}$ in the molten (top-left) and solidified (bottom-right) states. The fragment of crystal structure
is shown by blue spheres inside the crystallized structure.}
\label{fg:Ni2K}
\end{figure}

We have performed MD simulations of Ni$_{2047}$ clusters.
The initial geometry of the cluster was chosen to resemble icosahedral symmetry. 
Then the particle was
exposed to the heating-cooling cycle with the heating/cooling rate of 1 K/ps in 
the temperature range from 400 to 1400 K.
From Fig.~\ref{fg:Ni2K} is is seen that the final structure of the crystallized 
particle (blue dots at low temperatures) has lower energy 
than the initial icosahedral structure.  The method to obtain molecular
structures with higher binding energy by exposing the heated system to the cooling temperature bath
is widely used technique in MD simulations and is known as simulated annealing~\cite{Kirkpatrick:1983zz}. 
The lowering of the nickel cluster's energy after resolidification form icosahedral structure happens due to the formation of 
regions with FCC symmetry in the relatively large nickel clusters~\cite{Qi:2001jt}. 
From Fig.~\ref{fg:Ni2K} it is also seen a prominent hysteresis in the melting and crystallisation phase transitions due to the finite 
speed of the heating and cooling rates.
Note that the width of the hysteresis is almost $400$ K for the heating/cooling rate of 1 K/ps, which
shows that the crystallization process in the system is associated with transfer 
over a relatively high free energy barrier.

In the following section we investigate the dynamics of the crystallisation transition.    

\subsection{Solidification process in Ni$_{2047}$ clusters}
\begin{figure}[h]
\includegraphics[width=0.95\textwidth,clip]{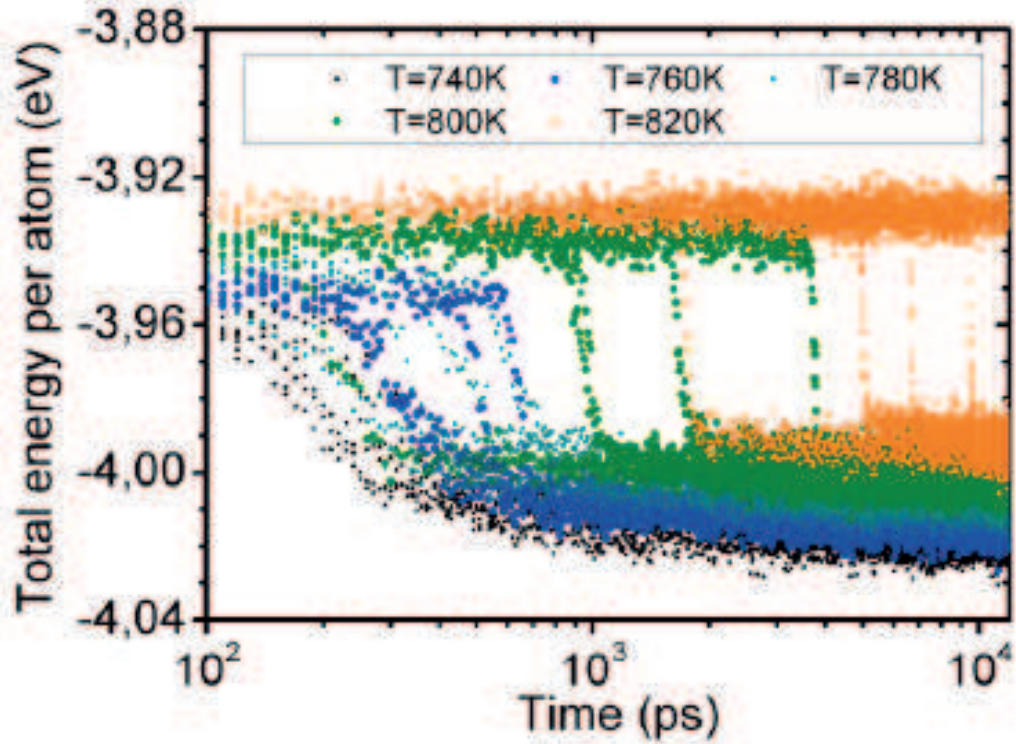}
\caption{Dependencies on time of the total energy per atom for the cluster being initially in the molten state at different temperatures of the
thermostat. Solidification phase transition occurs faster in the systems with lower temperature.}
\label{fg:Ni_cool_rates}
\end{figure}

By means of MD simulations we have investigated the solidification kinetics
of Ni$_{2047}$ cluster as a function of the amount of overcooling.
The initial structure
of the cluster was taken from MD simulations above the melting temperature. Then
the initially molten cluster was simulated at various temperatures below the melting point. We have
investigated the time needed for the system to change its phase from the molten to the solid state as a function of the
temperature below the phase transition point. The simulations were performed at 740~K, 760~K, 780~K, 800~K, 820~K, 830~K and 835~K.
For each temperature up to 15 independent simulations were produced. Fig.~\ref{fg:Ni_cool_rates} shows the dependence of the
total energy of the system on the simulation time for 5 randomly chosen trajectories for each value of temperature in the range between 740 and 820~K. Note the logarithm scale on the
horizontal axes.

From Fig.~\ref{fg:Ni_cool_rates} it is seen that at certain moments of time the energy of the system abruptly changes.
These moments correspond to the transition of the system from the molten to the solid state. Indeed, the solid state
of the cluster corresponds to the increased binding energy between the atoms due to the formation of the regular crystal lattice,
which leads to the lowering of the total energy of the system.

\subsection{Radial distribution function as an indicator of the phase transition}

In this section we analyse to which extent the radial distribution function (RDF) for atoms in the cluster
becomes affected by the solidification phase transition. RDF is defined as:

\begin{equation}
\label{eq:Rdist}
g(r)= \rho (r) / \bar{\rho} ,
\end{equation}
\noindent where $\rho(r)$ is the density of atoms at distance $r$ form a reference particle and $\bar\rho$ is an average density. $g(r)$ characterises the number of particles at the certain radial distance from a reference particle.

\begin{figure}[h]
\includegraphics[width=0.75\textwidth,clip]{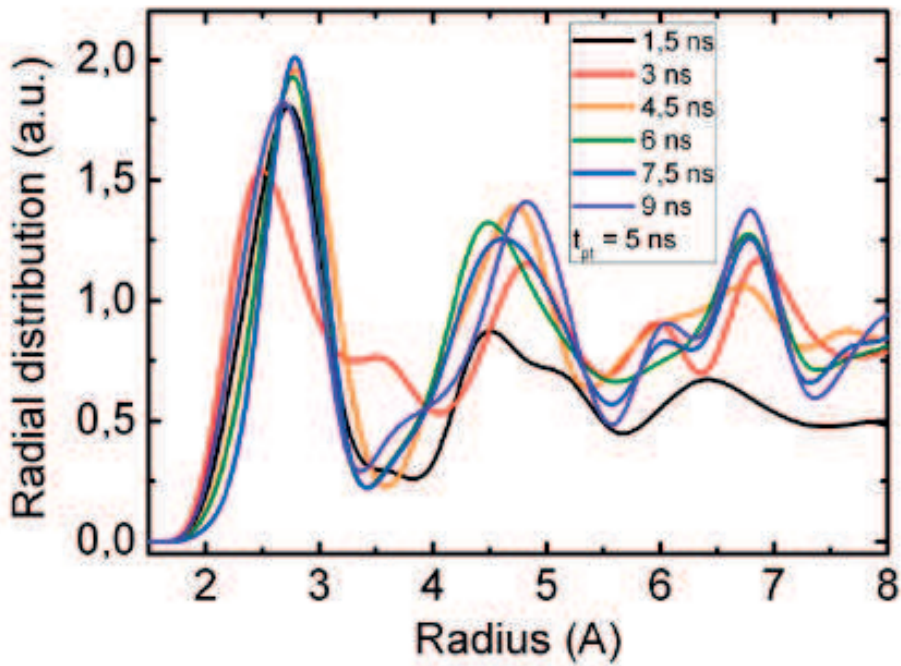}
\caption{Radial distribution function for the most central atoms of $Ni_{2047}$ cluster at different instances of simulation time.
For the selected for analysis trajectory the solidification transition had occurred at $t_{pt}=5$ ns.}
\label{fg:Rdist}
\end{figure}

We have calculated the RDF for nickel clusters in solid and molten states. 
The following procedure was adopted for the calculation of the RDF.
At certain instance of time the atoms located within 3~\AA~of the cluster center of mass were selected. For these atoms the
the RDF was calculated with a distance bin of 0.4~\AA~and averaged over 1 consequent nanosecond of simulation, which results
in averaging over 100 cluster structures, since we have written the structure of the system each 10 ps of simulation.
The RDF for the central atoms was calculated as an average over RDFs for
atoms of each of 100 structures.   

We analysed one particular trajectory
of the MD simulations at the thermostat temperature equal to 820 K (see orange trajectories in Fig.~\ref{fg:Ni_cool_rates}). 
For the chosen trajectory the solidification
transition occurs 5 ns after the start of the simulation. The RDF
for that trajectory at various instances of time is plotted using B-splines in Fig.~\ref{fg:Rdist}.

From Fig.~\ref{fg:Rdist} it is seen that RDFs at 1,5 and 3 ns (black and red curves) are
substantially different form those plotted for later time instances. This, obviously, can be explained
by the formation of cluster crystalline structure upon solidification. Indeed, peaks at $~5$ and $~7$~\AA~are more pronounced for the
RDFs at 6 $..$ 9 ns, since the crystalline structure has a long-range order. Note that RDF at 4,5 ns attain many  
features of RDFs characteristic for the crystallized state, however the crystallization transition takes place at 5 ns. This 
is related to the fact the calculated RDF was averaged over 1 ns in order to increase the statistics. Therefore,
RDF at 4,5 ns was partially constructed from the cluster structures being in crystalline state.

Despite that RDF can be used for the characterisation of the melting-solidification transition in the system, it is not entirely clear
how to define uniquely the phase transition moment from RDF analysis, especially when RDFs are calculated with a reduced
amount of sampling data, which is often the case for finite systems. In the next subsection we report on the analysis
of the diffusion coefficient behaviour which turns out to be 
much more convenient quantity for the characterization of the liquid-solid phase transition.
      
\subsection{Diffusion coefficient as a fingerprint of solidification phase transition}

The diffusion coefficient of a particle is defined as follows:

\begin{eqnarray}
\label{eq:diffusion}
D = \frac{\left\langle \Delta r^2 \right\rangle}{2 z \Delta t},
\end{eqnarray}
\noindent where $\left\langle \Delta r^2 \right\rangle$ is a
mean-square displacement of a particle per time $\Delta t$, and $z$ is dimensionality of space (3 for 3-dimensional diffusion)~\cite{Dick11}.

\begin{figure}[!h]
\includegraphics[width=0.95\textwidth,clip]{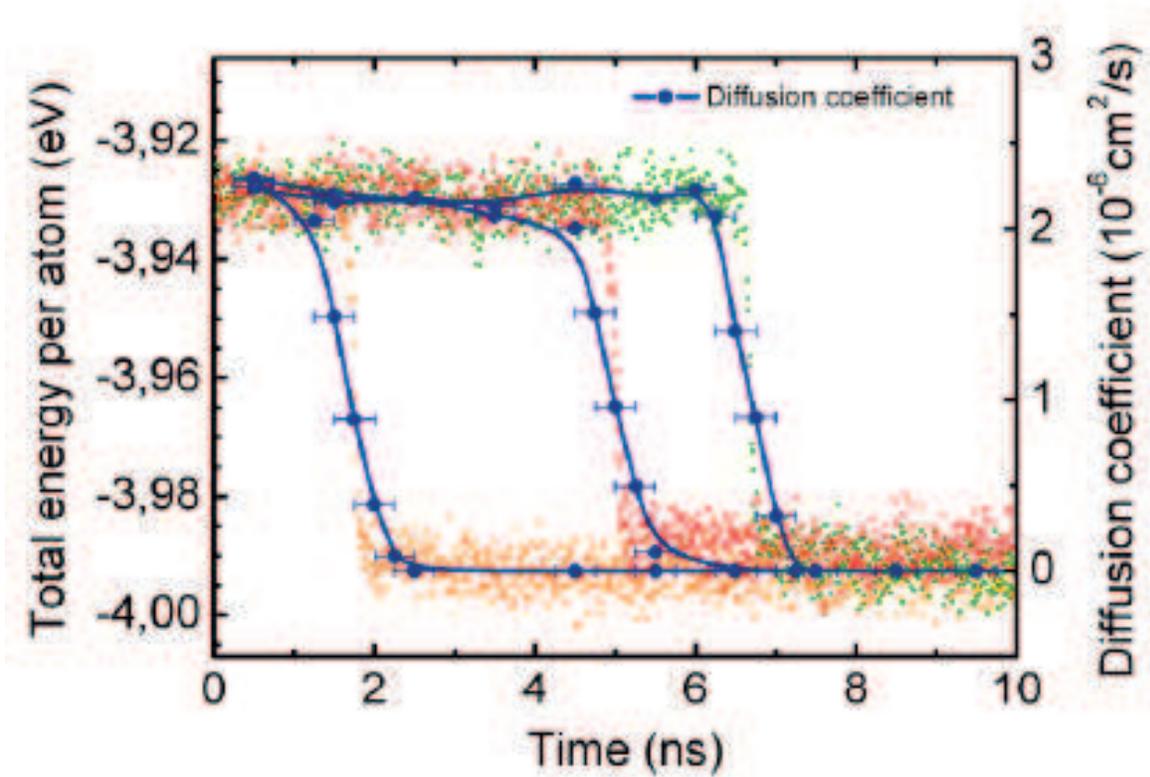}
\caption{Left axis: dependencies of the total energy per atom on time. Colours indicate
different MD trajectories. Right axis: dependence of diffusion coefficient
on time. }
\label{fg:Diffusion}
\end{figure}

We have calculated the self-diffusion coefficient of nickel atoms located in the central part of the cluster at different instances of time.
For the analysis we have chosen 3 out five MD trajectories of the $N_{2057}$ cluster conducted with a thermostat temperature equal to 820 K.
The total energy of the system and the diffusion coefficient as a functions of time are plotted for each trajectory in Fig.~\ref{fg:Diffusion}.

The following procedure was used for the calculation of the self-diffusion coefficient:
the part of the MD trajectory starting 0,5 ns before and ending 0,5 ns after each given instant of time was selected and the
coordinates of all atoms were recorded. For each structure cluster was translated and rotated as a whole
in order to minimize the root-mean-square displacement of atoms from a reference structure.
This procedure allows to exclude rotational and translational motion of the cluster when 
calculating the diffusion coefficient. Then the selected part of the trajectory was split into 10 segments and the
diffusion coefficient was calculated for all atoms located within 10~\AA~of the cluster center of mass using Eq.~(\ref{eq:diffusion}). The
final value of the diffusion coefficient was calculated as an average over all central atoms and 10 trajectory segments.
Since the diffusion coefficient was calculated as an average over 1 ns of the MD simulation,
in Fig.~\ref{fg:Diffusion} we show the corresponding horizontal error bars for each value of the diffusion coefficient.

From Fig.~\ref{fg:Diffusion} it is seen that for all 3 trajectories the diffusion coefficient drops abruptly at certain moments 
of time exactly corresponding to the moment of the solidification phase transition. The value of the diffusion
coefficient for the molten state is $\sim 2.3\times 10^{-6}$ cm$^2$/s. The diffusion coefficient in the solid
state appears to be smaller than $10^{-5}$ $..$ $10^{-6}$ cm$^2$/s. We were not able to evaluate it accurately for a given set
of data since it requires more than 5 ns to observe even single event of exchange of mutual positions for a pair of atoms
at 820~K.

From Fig.~\ref{fg:Diffusion} it is seen that the dependence of the diffusion coefficient on time
qualitatively reproduces well the time dependence of the total energy. One can use the diffusion coefficient as a reliable 
quantity to characterize
 the solidification phase transition in the system especially if calculation of the total energy of the system during
MD simulations is undesirable, which might be the case for certain GPU-based computer codes.

\subsection{Kinetics of the solidification phase transition}

In this subsection we evaluate thermodynamic characteristics of the solidification phase transition
on the basis of the analysis of kinetics of this process obtained from MD simulations.  

\begin{figure}
\includegraphics[width=1\textwidth,clip]{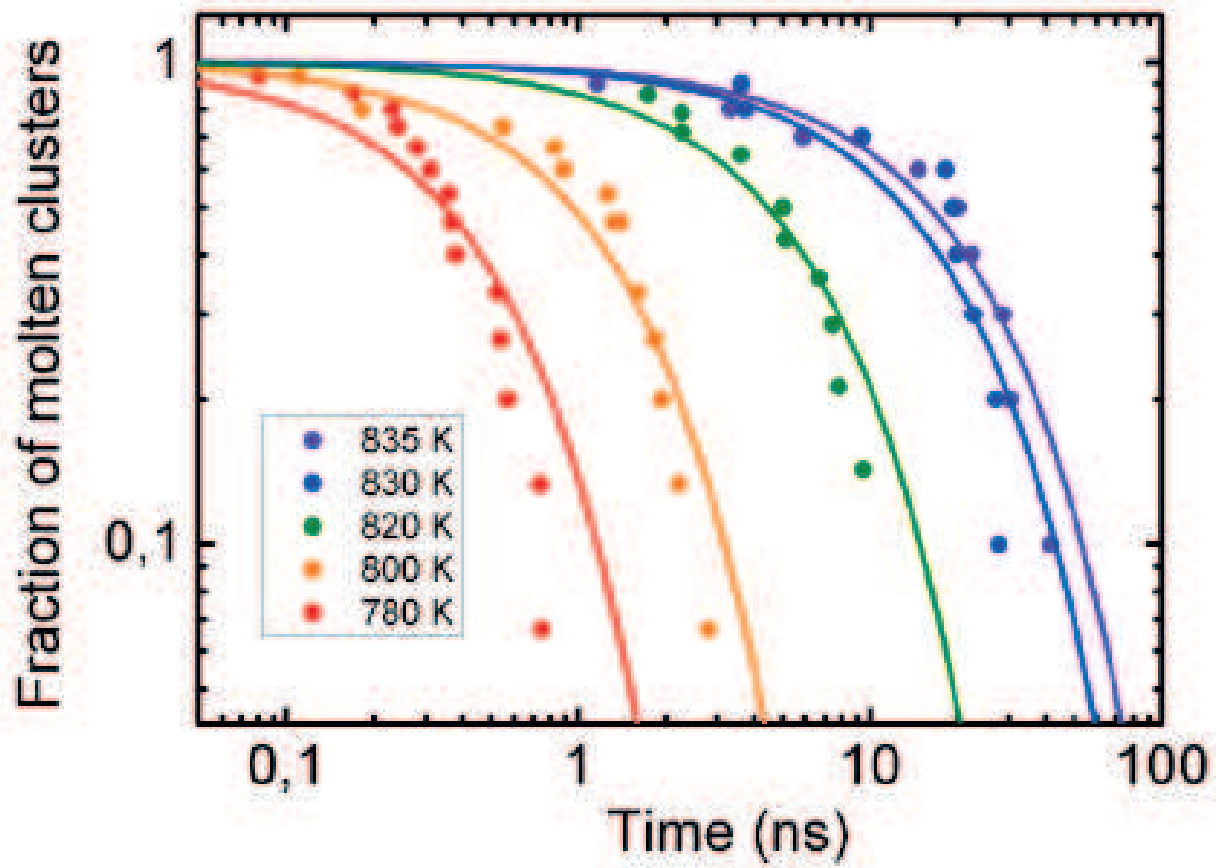}
\caption{Fraction of Ni$_{2057}$ clusters in the molten state as a function of simulation time for various temperatures below the phase transition point. Dots show the results obtained from MD simulations. Numerical fit of the results using Eq.~(\ref{eq:rate}) is shown by
solid lines.}
\label{fg:moltfrac}
\end{figure}

We calculate the latent heat of the phase transition as a difference
of the averaged total energy of the system in molten and solid states. For the Ni$_{2047}$ cluster the latent heat of the phase transition $\Delta E$ is
equal to 127.8 eV. The entropy density change between solid and molten states of the cluster can be calculated as follows:
\begin{eqnarray}
\label{eq:entdense}
\Delta s &=& \frac{\Delta E}{T_0 V_{cl}},
\end{eqnarray}
\noindent where T$_0$ and $V_{cl}$ is a phase transition temperature and cluster volume, correspondingly.
Note, that here we neglect the fact that the surface layer of the atoms of the cluster remains liquid after
the solidification of the cluster core. 


In Table~\ref{tab:LT} we present the life time of the system in overcooled state for temperature range between
780~K and 835~K. For our analysis we do not take into account simulations performed for temperatures lower than 780~K since the lifetime of the clusters in the overcooled state for that temperatures is comparable with the time necessary for the system to exchange
its energy with the thermostat in the course of the solidification process (see Fig.~\ref{fg:Ni_cool_rates}). 
The time of the phase transition was defined as a moment at which the energy of the system starts to decrease rapidly
due to the formation of the solid phase. The last row in Table~\ref{tab:LT} shows the escape rate from the overcooled states, which is defined as:

\begin{eqnarray}
\label{eq:rate2}
\kappa &=& \left( <t> \right)^{-1},
\end{eqnarray}
\noindent where $<t>$ denotes averaging of the lifetime over the simulation runs for the corresponding temperature.

\begin{figure}
\includegraphics[width=1\textwidth,clip]{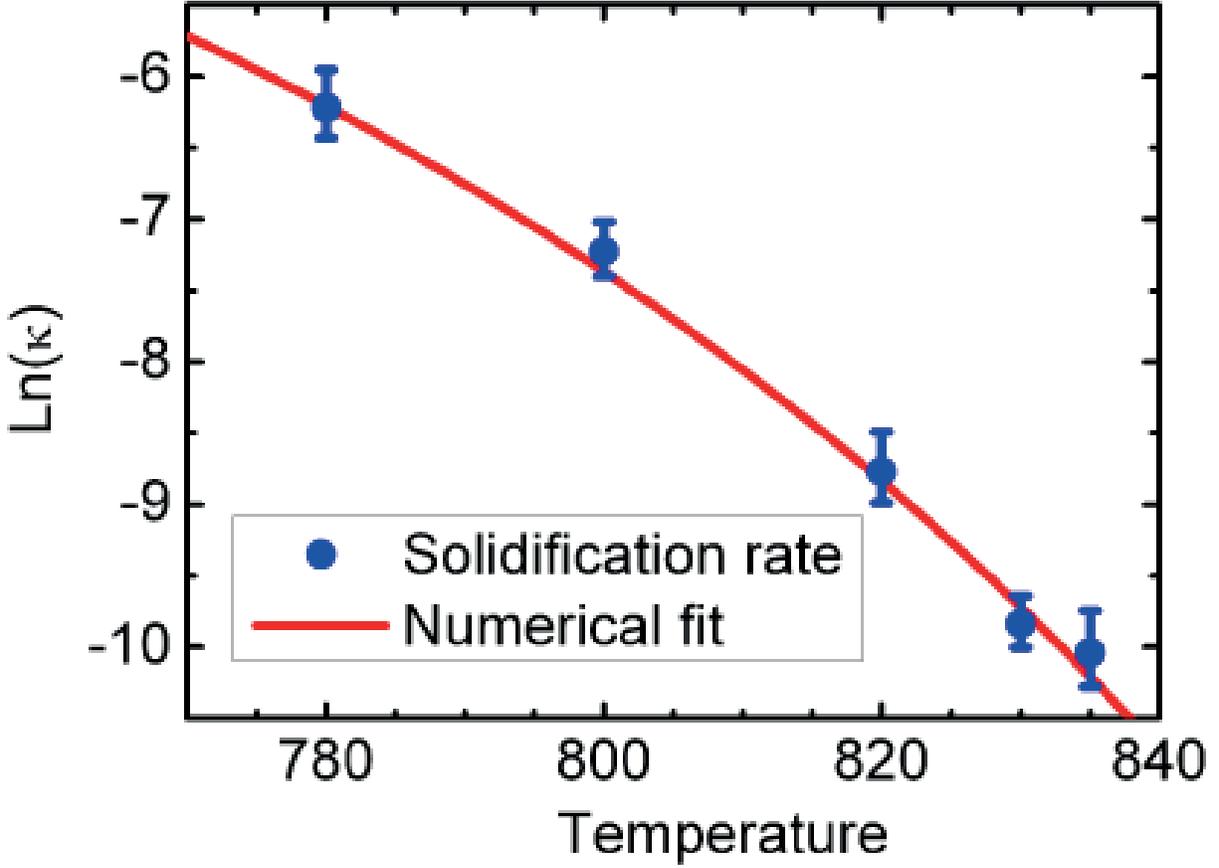}
\caption{Solidification rate of Ni$_{2057}$ cluster for various temperatures below the phase transition point. 
Error bars correspond to a single standard deviation calculated for a set of data for each temperature value. Numerical fit of the results using Eq.~(\ref{eq:rate3}) is shown by solid line. }
\label{fg:solrate}
\end{figure}

The fraction of clusters in the liquid state below the phase transition temperature as a function of time, $c(t)$, can be written as follows:

\begin{eqnarray}
\label{eq:exp_decay}
c(t) &=& \exp\left(-\kappa t \right),
\end{eqnarray}
\noindent where $\kappa$ can be evaluated using Eq.~(\ref{eq:rate2}). In Fig.\ref{fg:solrate} the results of the MD simulations
are shown by dots. Solid lines correspond to the exponential reduction of the fraction of liquid clusters as derived using Eq.~(\ref{eq:exp_decay}).

\begin{table}[htb]
\begin{center}
\begin{tabular}{| c | c | c | c | c | c | c  }
\hline
  & 780 K & 800 K & 820 K & 830 K & 835 K  \\
\hline
  1 & 80 & 110 & 100 & 3610 & 1160 \\
  2 & 170 & 170 & 1740 & 3730 & 3320 \\
  3 & 230 & 180 & 2260 & 9340 & 5890 \\
  4 & 240 & 550 & 3600 & 18130 & 14680 \\
  5 & 280 & 820 & 5020 & 19080 & 20008 \\
  6 & 310 & 890 & 5130 & 19760 & 22240 \\
  7 & 360 & 1260 & 6670 & 22640 & 28640 \\
  8 & 370 & 1320 & 7430 & 26670 & 30230 \\
  9 & 380 & 1390 & 7800 & 27670 & 41670 \\
  10 & 530 & 1600 & 9450 & 36380 & 63180 \\
  11 & 540 & 1830 & 11620 & -- & --  \\
  12 & 570 & 1930 & 23400 & -- & --  \\
  13 & 740 & 2210 & -- & -- & -- \\
  14 & 750 & 2800 & -- &--  &  --\\
  15 & 2000 & 3610 & -- & -- &  -- \\
\hline
mean (ps) & 503 & 1378 & 6440 & 18731 & 23102\\
$\kappa$ (10$^{-5}$ps$^{-1}$) & 199 & 72.6 & 15.5 & 5.34 & 4.33 \\
\hline
\end{tabular}
\end{center}
\caption{Life time in picoseconds of the overcooled liquid state for different MD trajectories at various temperatures. 15 simulations were performed for 780~K and 800~K, 12 for 820~K and 10 simulations for 830~K and 835~K.}
\label{tab:LT}
\end{table}

Knowing the dependence of the solidification rate of the clusters on temperature we can fit the results of MD simulations
to the Eq.~(\ref{eq:rate}) as follows:

\begin{eqnarray}
\label{eq:rate3}
\ln{\left(\kappa\cdot 1fs \right)} &=& a - \frac{b}{k_B T (T-T_0)^2},
\end{eqnarray}
\noindent where $a$, $b$ and $T_0$ are fitting parameters. The result of the fit of the MD simulations is shown in Fig.~\ref{fg:solrate},
and the fitting parameters have the following values: $a$=1.24, $b$=31053 eV$\cdot$K$^{2}$ and $T_0$=1029 K. Therefore, according to the results of the fitting, the precursor formation frequency (coefficient $A$ in Eq.~(\ref{eq:rate})) is 3.5 ps$^{-1}$ and phase transition temperature of the bulk nickel material for the used parametrization of the Sutton-Chen potential is equal to 1029 K. This value of phase transition temperature is lower than that, reported in~\cite{Shibuta10}, where the phase transition of the bulk nickel for the similar parametrization of the Sutton-Chen potential was calculated to be 1160 K. The discrepancies between our results and the results of previous calculation are largely because of limited statistics for the calculation of the overcooled clusters lifetimes.

Evaluating $\Delta s$ using Eq.~(\ref{eq:entdense}) one can calculate the liquid-solid surface tension coefficient $\alpha$ as a real root of the following equation:
\begin{eqnarray}
\label{eq:alpha}
\frac{16 \pi \alpha ^3}{3 \Delta s ^2} &=& b,
\end{eqnarray}
\noindent Solving Eq.~(\ref{eq:alpha}) one obtains $\alpha$ = 0.53 eV/nm$^2$ or 85 mJ/m$^2$. 
This value of $\alpha$ is approximately two times lower than that reported in Ref.~\cite{Qi:2001jt} for Nickel crystal-liquid
tension coefficient. The discrepancy can be attributed to the various parametrization of the Sutton-Chen potential used in our and that works. 

Knowing the surface tension coefficient $\alpha$ one can calculate the critical size of the precursor of the solid phase for different temperatures using Eq.~(\ref{eq:prec_crit}) and Eq.~(\ref{eq:entdense}). From that equations it follows that the critical precursor consists of 26, 33,
41, 53, 70, 80 and 87 atoms for
the temperature equal to 740, 760, 780, 800, 820, 830 and 835 K, correspondingly.


From Eq.~(\ref{eq:prec_crit}) it is seen that the critical size of the precursor grows when the temperature of the system approaches the
phase transition temperature of the bulk. However, for finite systems the size of the critical precursor can not be larger than the cluster size.
We can use
this idea to calculate the phase transition temperature of the finite size cluster systems as follows:

\begin{eqnarray}
\label{eq:t_melt_fs}
T(N)=T_0-\left(\frac{4 \pi \rho}{3 N} \right)^{1/3}\frac{2 \alpha}{\Delta s},
\end{eqnarray}
\noindent where $N$ denotes the number of particles in the cluster and $\rho$ is the particle density. This equation is also known as Gibbs-Thomson equation describing the
suppression of the melting point of the solid particles in their own fluid. From Eq.~(\ref{eq:t_melt_fs}) it follows that
the suppression of the melting point is inversely proportional to the number of atoms in the cluster in the power of one third,
and the melting point of the cluster consisting of 2057 atoms is equal to 962 K.

\section{Conclusions}
\label{sec:con}
In this paper we 
have conducted classical MD simulations of the Ni$_{2047}$ cluster with the use 
of many-body Sutton-Chen potential on the time scales up to 65 ns. 
For the purposes of this work we have developed an efficient software code
capable of performing large-scale MD simulations on GPUs.
We have demonstrated that with the use of GPU technology
one can achieve the speed-up of computations up to 400 times as compared to single core CPU.
On the basis of MD simulations we have investigated the radial distribution function and the diffusion coefficient
at molten and solidified states of the system.
We have analysed the solidification kinetics
of the clusters as a function of the over-cooling temperature and shown
that the kinetics of the phase transition can be described within the framework
of precursor formation theoretical model. Based on that theoretical model
we had derived various characteristics of the systems such
as solid-liquid surface tension coefficient, rate of the precursors formation,
phase transition temperature for the clusters of arbitrary size.
This work highlights the computational advantages one can achieve with
the use of GPUs and demonstrates a recipe to evaluate 
various thermodynamic characteristics of the finite system from 
a single set of MD simulations data. Further work can be devoted to the optimisation of the GPU code,
and investigating the large systems consisting of mixture of different metals (nanoalloys). With minor
modifications the code can be adopted for large-scale simulations of nanoindentation processes
in various metal alloys and investigation of dislocation dynamics.

\acknowledgments{The authors acknowledge Frankfurt Center for Scientific Computing
for the possibility to perform complex computer simulation using CPU and GPU. A.V.Y.
thanks Stiftung Polytechnische Gesellschaft Frankfurt am Main for financial support.


\begin{thebibliography}{53}%
\makeatletter
\providecommand \@ifxundefined [1]{%
 \@ifx{#1\undefined}
}%
\providecommand \@ifnum [1]{%
 \ifnum #1\expandafter \@firstoftwo
 \else \expandafter \@secondoftwo
 \fi
}%
\providecommand \@ifx [1]{%
 \ifx #1\expandafter \@firstoftwo
 \else \expandafter \@secondoftwo
 \fi
}%
\providecommand \natexlab [1]{#1}%
\providecommand \enquote  [1]{``#1''}%
\providecommand \bibnamefont  [1]{#1}%
\providecommand \bibfnamefont [1]{#1}%
\providecommand \citenamefont [1]{#1}%
\providecommand \href@noop [0]{\@secondoftwo}%
\providecommand \href [0]{\begingroup \@sanitize@url \@href}%
\providecommand \@href[1]{\@@startlink{#1}\@@href}%
\providecommand \@@href[1]{\endgroup#1\@@endlink}%
\providecommand \@sanitize@url [0]{\catcode `\\12\catcode `\$12\catcode
  `\&12\catcode `\#12\catcode `\^12\catcode `\_12\catcode `\%12\relax}%
\providecommand \@@startlink[1]{}%
\providecommand \@@endlink[0]{}%
\providecommand \url  [0]{\begingroup\@sanitize@url \@url }%
\providecommand \@url [1]{\endgroup\@href {#1}{\urlprefix }}%
\providecommand \urlprefix  [0]{URL }%
\providecommand \Eprint [0]{\href }%
\providecommand \doibase [0]{http://dx.doi.org/}%
\providecommand \selectlanguage [0]{\@gobble}%
\providecommand \bibinfo  [0]{\@secondoftwo}%
\providecommand \bibfield  [0]{\@secondoftwo}%
\providecommand \translation [1]{[#1]}%
\providecommand \BibitemOpen [0]{}%
\providecommand \bibitemStop [0]{}%
\providecommand \bibitemNoStop [0]{.\EOS\space}%
\providecommand \EOS [0]{\spacefactor3000\relax}%
\providecommand \BibitemShut  [1]{\csname bibitem#1\endcsname}%
\let\auto@bib@innerbib\@empty
\bibitem [{\citenamefont {Aguado}\ and\ \citenamefont
  {Jarrold}(2011)}]{Aguado11}%
  \BibitemOpen
  \bibfield  {author} {\bibinfo {author} {\bibfnamefont {A.}~\bibnamefont
  {Aguado}}\ and\ \bibinfo {author} {\bibfnamefont {M.}~\bibnamefont
  {Jarrold}},\ }\href@noop {} {\bibfield  {journal} {\bibinfo  {journal}
  {Annu.\ Rev.\ Phys.\ Chem.}\ }\textbf {\bibinfo {volume} {62}},\ \bibinfo
  {pages} {151} (\bibinfo {year} {2011})}\BibitemShut {NoStop}%
\bibitem [{\citenamefont {Turnbull}(1950)}]{Turnbull50}%
  \BibitemOpen
  \bibfield  {author} {\bibinfo {author} {\bibfnamefont {D.}~\bibnamefont
  {Turnbull}},\ }\href@noop {} {\bibfield  {journal} {\bibinfo  {journal} {J.
  Appl. Phys.}\ }\textbf {\bibinfo {volume} {21}},\ \bibinfo {pages} {1022}
  (\bibinfo {year} {1950})}\BibitemShut {NoStop}%
\bibitem [{\citenamefont {Turnbull}(1952)}]{Turnbull52}%
  \BibitemOpen
  \bibfield  {author} {\bibinfo {author} {\bibfnamefont {D.}~\bibnamefont
  {Turnbull}},\ }\href@noop {} {\bibfield  {journal} {\bibinfo  {journal} {J.
  Comp. Phys.}\ }\textbf {\bibinfo {volume} {20}},\ \bibinfo {pages} {411}
  (\bibinfo {year} {1952})}\BibitemShut {NoStop}%
\bibitem [{\citenamefont {Pawlow}(1909)}]{Pawlow1909}%
  \BibitemOpen
  \bibfield  {author} {\bibinfo {author} {\bibfnamefont {P.}~\bibnamefont
  {Pawlow}},\ }\href@noop {} {\bibfield  {journal} {\bibinfo  {journal} {Z.
  Phys. Chem.}\ }\textbf {\bibinfo {volume} {65}},\ \bibinfo {pages} {545}
  (\bibinfo {year} {1909})}\BibitemShut {NoStop}%
\bibitem [{\citenamefont {Buffat}\ and\ \citenamefont
  {Borel}(1976)}]{Buffat76}%
  \BibitemOpen
  \bibfield  {author} {\bibinfo {author} {\bibfnamefont {P.}~\bibnamefont
  {Buffat}}\ and\ \bibinfo {author} {\bibfnamefont {J.-P.}\ \bibnamefont
  {Borel}},\ }\href@noop {} {\bibfield  {journal} {\bibinfo  {journal} {Phys.
  Rev. A}\ }\textbf {\bibinfo {volume} {13}},\ \bibinfo {pages} {2287}
  (\bibinfo {year} {1976})}\BibitemShut {NoStop}%
\bibitem [{\citenamefont {Castro}\ \emph {et~al.}(1990)\citenamefont {Castro},
  \citenamefont {Reifenberger}, \citenamefont {Choi},\ and\ \citenamefont
  {Anders}}]{Castro90}%
  \BibitemOpen
  \bibfield  {author} {\bibinfo {author} {\bibfnamefont {T.}~\bibnamefont
  {Castro}}, \bibinfo {author} {\bibfnamefont {R.}~\bibnamefont
  {Reifenberger}}, \bibinfo {author} {\bibfnamefont {E.}~\bibnamefont {Choi}},
  \ and\ \bibinfo {author} {\bibfnamefont {R.}~\bibnamefont {Anders}},\
  }\href@noop {} {\bibfield  {journal} {\bibinfo  {journal} {Phys. Rev. B}\
  }\textbf {\bibinfo {volume} {42}},\ \bibinfo {pages} {8548} (\bibinfo {year}
  {1990})}\BibitemShut {NoStop}%
\bibitem [{\citenamefont {Lai}\ \emph {et~al.}(1996)\citenamefont {Lai},
  \citenamefont {Guo}, \citenamefont {Petrova}, \citenamefont {Ramanath},\ and\
  \citenamefont {Allen}}]{Lai96}%
  \BibitemOpen
  \bibfield  {author} {\bibinfo {author} {\bibfnamefont {S.}~\bibnamefont
  {Lai}}, \bibinfo {author} {\bibfnamefont {J.}~\bibnamefont {Guo}}, \bibinfo
  {author} {\bibfnamefont {V.}~\bibnamefont {Petrova}}, \bibinfo {author}
  {\bibfnamefont {G.}~\bibnamefont {Ramanath}}, \ and\ \bibinfo {author}
  {\bibfnamefont {L.}~\bibnamefont {Allen}},\ }\href@noop {} {\bibfield
  {journal} {\bibinfo  {journal} {Phys. Rev. Lett.}\ }\textbf {\bibinfo
  {volume} {77}},\ \bibinfo {pages} {99} (\bibinfo {year} {1996})}\BibitemShut
  {NoStop}%
\bibitem [{\citenamefont {Bottani}\ \emph {et~al.}(1999)\citenamefont
  {Bottani}, \citenamefont {Bassi},\ and\ \citenamefont {Tanner}}]{Bottani90}%
  \BibitemOpen
  \bibfield  {author} {\bibinfo {author} {\bibfnamefont {C.}~\bibnamefont
  {Bottani}}, \bibinfo {author} {\bibfnamefont {A.}~\bibnamefont {Bassi}}, \
  and\ \bibinfo {author} {\bibfnamefont {B.}~\bibnamefont {Tanner}},\
  }\href@noop {} {\bibfield  {journal} {\bibinfo  {journal} {Phys. Rev. B}\
  }\textbf {\bibinfo {volume} {59}},\ \bibinfo {pages} {R15601} (\bibinfo
  {year} {1999})}\BibitemShut {NoStop}%
\bibitem [{\citenamefont {Schmidt}\ \emph {et~al.}(1997)\citenamefont
  {Schmidt}, \citenamefont {Kusche}, \citenamefont {Kronm{\"u}ller},
  \citenamefont {von Issendorff},\ and\ \citenamefont
  {Haberland}}]{Haberland97}%
  \BibitemOpen
  \bibfield  {author} {\bibinfo {author} {\bibfnamefont {M.}~\bibnamefont
  {Schmidt}}, \bibinfo {author} {\bibfnamefont {R.}~\bibnamefont {Kusche}},
  \bibinfo {author} {\bibfnamefont {W.}~\bibnamefont {Kronm{\"u}ller}},
  \bibinfo {author} {\bibfnamefont {B.}~\bibnamefont {von Issendorff}}, \ and\
  \bibinfo {author} {\bibfnamefont {H.}~\bibnamefont {Haberland}},\ }\href@noop
  {} {\bibfield  {journal} {\bibinfo  {journal} {Phys. Rev. Lett.}\ }\textbf
  {\bibinfo {volume} {79}},\ \bibinfo {pages} {99} (\bibinfo {year}
  {1997})}\BibitemShut {NoStop}%
\bibitem [{\citenamefont {Schmidt}\ \emph {et~al.}(1998)\citenamefont
  {Schmidt}, \citenamefont {Kusche}, \citenamefont {von Issendorff},\ and\
  \citenamefont {Haberland}}]{Haberland98}%
  \BibitemOpen
  \bibfield  {author} {\bibinfo {author} {\bibfnamefont {M.}~\bibnamefont
  {Schmidt}}, \bibinfo {author} {\bibfnamefont {R.}~\bibnamefont {Kusche}},
  \bibinfo {author} {\bibfnamefont {B.}~\bibnamefont {von Issendorff}}, \ and\
  \bibinfo {author} {\bibfnamefont {H.}~\bibnamefont {Haberland}},\ }\href@noop
  {} {\bibfield  {journal} {\bibinfo  {journal} {Nature}\ }\textbf {\bibinfo
  {volume} {393}},\ \bibinfo {pages} {238} (\bibinfo {year}
  {1998})}\BibitemShut {NoStop}%
\bibitem [{\citenamefont {Kusche}\ \emph {et~al.}(1999)\citenamefont {Kusche},
  \citenamefont {Hippler}, \citenamefont {Schmidt}, \citenamefont {von
  Issendorff},\ and\ \citenamefont {Haberland}}]{Haberland99}%
  \BibitemOpen
  \bibfield  {author} {\bibinfo {author} {\bibfnamefont {R.}~\bibnamefont
  {Kusche}}, \bibinfo {author} {\bibfnamefont {T.}~\bibnamefont {Hippler}},
  \bibinfo {author} {\bibfnamefont {M.}~\bibnamefont {Schmidt}}, \bibinfo
  {author} {\bibfnamefont {B.}~\bibnamefont {von Issendorff}}, \ and\ \bibinfo
  {author} {\bibfnamefont {H.}~\bibnamefont {Haberland}},\ }\href@noop {}
  {\bibfield  {journal} {\bibinfo  {journal} {Eur.\ Phys.\ J.\ D}\ }\textbf
  {\bibinfo {volume} {9}},\ \bibinfo {pages} {1} (\bibinfo {year}
  {1999})}\BibitemShut {NoStop}%
\bibitem [{\citenamefont {Haberland}\ \emph {et~al.}(1999)\citenamefont
  {Haberland}, \citenamefont {Hippler}, \citenamefont {Donges}, \citenamefont
  {Kostko}, \citenamefont {Schmidt}, \citenamefont {von Issendorff},\ and\
  \citenamefont {Haberland}}]{Haberland05}%
  \BibitemOpen
  \bibfield  {author} {\bibinfo {author} {\bibfnamefont {H.}~\bibnamefont
  {Haberland}}, \bibinfo {author} {\bibfnamefont {T.}~\bibnamefont {Hippler}},
  \bibinfo {author} {\bibfnamefont {J.}~\bibnamefont {Donges}}, \bibinfo
  {author} {\bibfnamefont {O.}~\bibnamefont {Kostko}}, \bibinfo {author}
  {\bibfnamefont {M.}~\bibnamefont {Schmidt}}, \bibinfo {author} {\bibfnamefont
  {B.}~\bibnamefont {von Issendorff}}, \ and\ \bibinfo {author} {\bibfnamefont
  {H.}~\bibnamefont {Haberland}},\ }\href@noop {} {\bibfield  {journal}
  {\bibinfo  {journal} {Eur.\ Phys.\ J.\ D}\ }\textbf {\bibinfo {volume} {9}},\
  \bibinfo {pages} {1} (\bibinfo {year} {1999})}\BibitemShut {NoStop}%
\bibitem [{\citenamefont {Calvo}\ and\ \citenamefont
  {Spiegelmann}(2006)}]{Calvo00}%
  \BibitemOpen
  \bibfield  {author} {\bibinfo {author} {\bibfnamefont {F.}~\bibnamefont
  {Calvo}}\ and\ \bibinfo {author} {\bibfnamefont {F.}~\bibnamefont
  {Spiegelmann}},\ }\href@noop {} {\bibfield  {journal} {\bibinfo  {journal}
  {J. Comp. Phys.}\ }\textbf {\bibinfo {volume} {112}},\ \bibinfo {pages}
  {2888} (\bibinfo {year} {2006})}\BibitemShut {NoStop}%
\bibitem [{\citenamefont {Reyes-Nava}\ \emph {et~al.}(2003)\citenamefont
  {Reyes-Nava}, \citenamefont {Garzon},\ and\ \citenamefont
  {Michaelian}}]{ReyesNava03}%
  \BibitemOpen
  \bibfield  {author} {\bibinfo {author} {\bibfnamefont {J.}~\bibnamefont
  {Reyes-Nava}}, \bibinfo {author} {\bibfnamefont {I.}~\bibnamefont {Garzon}},
  \ and\ \bibinfo {author} {\bibfnamefont {K.}~\bibnamefont {Michaelian}},\
  }\href@noop {} {\bibfield  {journal} {\bibinfo  {journal} {Phys. Rev. B}\
  }\textbf {\bibinfo {volume} {67}},\ \bibinfo {pages} {165401} (\bibinfo
  {year} {2003})}\BibitemShut {NoStop}%
\bibitem [{\citenamefont {Calvo}\ and\ \citenamefont
  {Spiegelmann}(2004)}]{Calvo04}%
  \BibitemOpen
  \bibfield  {author} {\bibinfo {author} {\bibfnamefont {F.}~\bibnamefont
  {Calvo}}\ and\ \bibinfo {author} {\bibfnamefont {F.}~\bibnamefont
  {Spiegelmann}},\ }\href@noop {} {\bibfield  {journal} {\bibinfo  {journal}
  {J. Comp. Phys.}\ }\textbf {\bibinfo {volume} {120}},\ \bibinfo {pages}
  {9684} (\bibinfo {year} {2004})}\BibitemShut {NoStop}%
\bibitem [{\citenamefont {Manninen}\ \emph {et~al.}(2004)\citenamefont
  {Manninen}, \citenamefont {Rytkonen},\ and\ \citenamefont
  {Manninen}}]{Manninen04}%
  \BibitemOpen
  \bibfield  {author} {\bibinfo {author} {\bibfnamefont {K.}~\bibnamefont
  {Manninen}}, \bibinfo {author} {\bibfnamefont {A.}~\bibnamefont {Rytkonen}},
  \ and\ \bibinfo {author} {\bibfnamefont {M.}~\bibnamefont {Manninen}},\
  }\href@noop {} {\bibfield  {journal} {\bibinfo  {journal} {Eur.\ Phys.\ J.\
  D}\ }\textbf {\bibinfo {volume} {29}},\ \bibinfo {pages} {39} (\bibinfo
  {year} {2004})}\BibitemShut {NoStop}%
\bibitem [{\citenamefont {Chacko}\ \emph {et~al.}(2005)\citenamefont {Chacko},
  \citenamefont {Kanhere},\ and\ \citenamefont {Blundell}}]{Chacko05}%
  \BibitemOpen
  \bibfield  {author} {\bibinfo {author} {\bibfnamefont {S.}~\bibnamefont
  {Chacko}}, \bibinfo {author} {\bibfnamefont {D.}~\bibnamefont {Kanhere}}, \
  and\ \bibinfo {author} {\bibfnamefont {S.}~\bibnamefont {Blundell}},\
  }\href@noop {} {\bibfield  {journal} {\bibinfo  {journal} {Phys. Rev. B}\
  }\textbf {\bibinfo {volume} {71}},\ \bibinfo {pages} {155407} (\bibinfo
  {year} {2005})}\BibitemShut {NoStop}%
\bibitem [{\citenamefont {Aguado}\ and\ \citenamefont
  {L{\'o}pez}(2005)}]{Aguado05a}%
  \BibitemOpen
  \bibfield  {author} {\bibinfo {author} {\bibfnamefont {A.}~\bibnamefont
  {Aguado}}\ and\ \bibinfo {author} {\bibfnamefont {J.}~\bibnamefont
  {L{\'o}pez}},\ }\href@noop {} {\bibfield  {journal} {\bibinfo  {journal}
  {Phys. Rev. Lett.}\ }\textbf {\bibinfo {volume} {94}},\ \bibinfo {pages}
  {233401} (\bibinfo {year} {2005})}\BibitemShut {NoStop}%
\bibitem [{\citenamefont {Aguado}(2005)}]{Aguado05b}%
  \BibitemOpen
  \bibfield  {author} {\bibinfo {author} {\bibfnamefont {A.}~\bibnamefont
  {Aguado}},\ }\href@noop {} {\bibfield  {journal} {\bibinfo  {journal}
  {J.~Phys.\ Chem.~B}\ }\textbf {\bibinfo {volume} {109}},\ \bibinfo {pages}
  {13043} (\bibinfo {year} {2005})}\BibitemShut {NoStop}%
\bibitem [{\citenamefont {Noya}\ \emph {et~al.}(2007)\citenamefont {Noya},
  \citenamefont {Doye}, \citenamefont {Wales},\ and\ \citenamefont
  {Aguado}}]{Noya07}%
  \BibitemOpen
  \bibfield  {author} {\bibinfo {author} {\bibfnamefont {E.}~\bibnamefont
  {Noya}}, \bibinfo {author} {\bibfnamefont {J.}~\bibnamefont {Doye}}, \bibinfo
  {author} {\bibfnamefont {D.}~\bibnamefont {Wales}}, \ and\ \bibinfo {author}
  {\bibfnamefont {A.}~\bibnamefont {Aguado}},\ }\href@noop {} {\bibfield
  {journal} {\bibinfo  {journal} {Eur.\ Phys.\ J.\ D}\ }\textbf {\bibinfo
  {volume} {43}},\ \bibinfo {pages} {57} (\bibinfo {year} {2007})}\BibitemShut
  {NoStop}%
\bibitem [{\citenamefont {Shvartsburg}\ and\ \citenamefont
  {Jarrold}(2000)}]{Shvartsburg00}%
  \BibitemOpen
  \bibfield  {author} {\bibinfo {author} {\bibfnamefont {A.}~\bibnamefont
  {Shvartsburg}}\ and\ \bibinfo {author} {\bibfnamefont {M.}~\bibnamefont
  {Jarrold}},\ }\href@noop {} {\bibfield  {journal} {\bibinfo  {journal} {Phys.
  Rev. Lett.}\ }\textbf {\bibinfo {volume} {85}},\ \bibinfo {pages} {2530}
  (\bibinfo {year} {2000})}\BibitemShut {NoStop}%
\bibitem [{\citenamefont {Breaux}\ \emph {et~al.}(2003)\citenamefont {Breaux},
  \citenamefont {Benirschke}, \citenamefont {Sugai}, \citenamefont {Kinnear},\
  and\ \citenamefont {Jarrold}}]{Breaux03}%
  \BibitemOpen
  \bibfield  {author} {\bibinfo {author} {\bibfnamefont {G.}~\bibnamefont
  {Breaux}}, \bibinfo {author} {\bibfnamefont {R.}~\bibnamefont {Benirschke}},
  \bibinfo {author} {\bibfnamefont {T.}~\bibnamefont {Sugai}}, \bibinfo
  {author} {\bibfnamefont {B.}~\bibnamefont {Kinnear}}, \ and\ \bibinfo
  {author} {\bibfnamefont {M.}~\bibnamefont {Jarrold}},\ }\href@noop {}
  {\bibfield  {journal} {\bibinfo  {journal} {Phys. Rev. Lett.}\ }\textbf
  {\bibinfo {volume} {91}},\ \bibinfo {pages} {215508} (\bibinfo {year}
  {2003})}\BibitemShut {NoStop}%
\bibitem [{\citenamefont {Lyalin}\ \emph {et~al.}(2009)\citenamefont {Lyalin},
  \citenamefont {Hussien}, \citenamefont {Solov’yov},\ and\ \citenamefont
  {Greiner}}]{Lyalin09}%
  \BibitemOpen
  \bibfield  {author} {\bibinfo {author} {\bibfnamefont {A.}~\bibnamefont
  {Lyalin}}, \bibinfo {author} {\bibfnamefont {A.}~\bibnamefont {Hussien}},
  \bibinfo {author} {\bibfnamefont {A.}~\bibnamefont {Solov’yov}}, \ and\
  \bibinfo {author} {\bibfnamefont {W.}~\bibnamefont {Greiner}},\ }\href@noop
  {} {\bibfield  {journal} {\bibinfo  {journal} {Phys. Rev. B}\ }\textbf
  {\bibinfo {volume} {79}},\ \bibinfo {pages} {165403} (\bibinfo {year}
  {2009})}\BibitemShut {NoStop}%
\bibitem [{\citenamefont {Ding}\ \emph {et~al.}(2004)\citenamefont {Ding},
  \citenamefont {Bolton},\ and\ \citenamefont {Rosén}}]{Ding04}%
  \BibitemOpen
  \bibfield  {author} {\bibinfo {author} {\bibfnamefont {F.}~\bibnamefont
  {Ding}}, \bibinfo {author} {\bibfnamefont {K.}~\bibnamefont {Bolton}}, \ and\
  \bibinfo {author} {\bibfnamefont {A.}~\bibnamefont {Rosén}},\ }\href@noop {}
  {\bibfield  {journal} {\bibinfo  {journal} {J. Vac. Sci. Technol. A}\
  }\textbf {\bibinfo {volume} {22}},\ \bibinfo {pages} {1471} (\bibinfo {year}
  {2004})}\BibitemShut {NoStop}%
\bibitem [{\citenamefont {Jiang}\ \emph {et~al.}(2007)\citenamefont {Jiang},
  \citenamefont {Awasthi}, \citenamefont {Kolmogorov}, \citenamefont
  {Setyawan}, \citenamefont {Börjesson}, \citenamefont {Bolton},
  \citenamefont {Harutyunyan},\ and\ \citenamefont {Curtarolo}}]{Cutarolo07}%
  \BibitemOpen
  \bibfield  {author} {\bibinfo {author} {\bibfnamefont {A.}~\bibnamefont
  {Jiang}}, \bibinfo {author} {\bibfnamefont {N.}~\bibnamefont {Awasthi}},
  \bibinfo {author} {\bibfnamefont {A.}~\bibnamefont {Kolmogorov}}, \bibinfo
  {author} {\bibfnamefont {W.}~\bibnamefont {Setyawan}}, \bibinfo {author}
  {\bibfnamefont {A.}~\bibnamefont {Börjesson}}, \bibinfo {author}
  {\bibfnamefont {K.}~\bibnamefont {Bolton}}, \bibinfo {author} {\bibfnamefont
  {A.}~\bibnamefont {Harutyunyan}}, \ and\ \bibinfo {author} {\bibfnamefont
  {S.}~\bibnamefont {Curtarolo}},\ }\href@noop {} {\bibfield  {journal}
  {\bibinfo  {journal} {Phys. Rev. B}\ }\textbf {\bibinfo {volume} {75}},\
  \bibinfo {pages} {205426} (\bibinfo {year} {2007})}\BibitemShut {NoStop}%
\bibitem [{\citenamefont {Petrov}\ and\ \citenamefont
  {Fur{\'o}}(2006)}]{Petrov:2006gf}%
  \BibitemOpen
  \bibfield  {author} {\bibinfo {author} {\bibfnamefont {O.}~\bibnamefont
  {Petrov}}\ and\ \bibinfo {author} {\bibfnamefont {I.}~\bibnamefont
  {Fur{\'o}}},\ }\href@noop {} {\bibfield  {journal} {\bibinfo  {journal}
  {Phys. Rev. E}\ }\textbf {\bibinfo {volume} {73}} (\bibinfo {year}
  {2006})}\BibitemShut {NoStop}%
\bibitem [{\citenamefont {Ruban}\ and\ \citenamefont
  {Abrikosov}(2008)}]{Ruban:2008ko}%
  \BibitemOpen
  \bibfield  {author} {\bibinfo {author} {\bibfnamefont {A.~V.}\ \bibnamefont
  {Ruban}}\ and\ \bibinfo {author} {\bibfnamefont {I.~A.}\ \bibnamefont
  {Abrikosov}},\ }\href@noop {} {\bibfield  {journal} {\bibinfo  {journal}
  {Reports on Progress in Physics}\ }\textbf {\bibinfo {volume} {71}},\
  \bibinfo {pages} {046501} (\bibinfo {year} {2008})}\BibitemShut {NoStop}%
\bibitem [{\citenamefont {Aguado}\ \emph {et~al.}(2001)\citenamefont {Aguado},
  \citenamefont {L{\'o}pez}, \citenamefont {Alonso},\ and\ \citenamefont
  {Stott}}]{Aguado:2001ie}%
  \BibitemOpen
  \bibfield  {author} {\bibinfo {author} {\bibfnamefont {A.}~\bibnamefont
  {Aguado}}, \bibinfo {author} {\bibfnamefont {J.~M.}\ \bibnamefont
  {L{\'o}pez}}, \bibinfo {author} {\bibfnamefont {J.~A.}\ \bibnamefont
  {Alonso}}, \ and\ \bibinfo {author} {\bibfnamefont {M.~J.}\ \bibnamefont
  {Stott}},\ }\href@noop {} {\bibfield  {journal} {\bibinfo  {journal}
  {J.~Phys.\ Chem.~B}\ }\textbf {\bibinfo {volume} {105}},\ \bibinfo {pages}
  {2386} (\bibinfo {year} {2001})}\BibitemShut {NoStop}%
\bibitem [{\citenamefont {Goringe}\ \emph {et~al.}(1999)\citenamefont
  {Goringe}, \citenamefont {Bowler},\ and\ \citenamefont
  {Hern{\'a}ndez}}]{Goringe:1999bx}%
  \BibitemOpen
  \bibfield  {author} {\bibinfo {author} {\bibfnamefont {C.~M.}\ \bibnamefont
  {Goringe}}, \bibinfo {author} {\bibfnamefont {D.~R.}\ \bibnamefont {Bowler}},
  \ and\ \bibinfo {author} {\bibfnamefont {E.}~\bibnamefont {Hern{\'a}ndez}},\
  }\href@noop {} {\bibfield  {journal} {\bibinfo  {journal} {Reports on
  Progress in Physics}\ }\textbf {\bibinfo {volume} {60}},\ \bibinfo {pages}
  {1447} (\bibinfo {year} {1999})}\BibitemShut {NoStop}%
\bibitem [{\citenamefont {Yue}\ \emph {et~al.}(2001)\citenamefont {Yue},
  \citenamefont {Cagin}, \citenamefont {Johnson},\ and\ \citenamefont
  {Goddard}}]{Qi:2001jt}%
  \BibitemOpen
  \bibfield  {author} {\bibinfo {author} {\bibfnamefont {Q.}~\bibnamefont
  {Yue}}, \bibinfo {author} {\bibfnamefont {T.}~\bibnamefont {Cagin}}, \bibinfo
  {author} {\bibfnamefont {W.}~\bibnamefont {Johnson}}, \ and\ \bibinfo
  {author} {\bibfnamefont {W.}~\bibnamefont {Goddard}},\ }\href@noop {}
  {\bibfield  {journal} {\bibinfo  {journal} {J. Comp. Phys.}\ }\textbf
  {\bibinfo {volume} {115}},\ \bibinfo {pages} {385} (\bibinfo {year}
  {2001})}\BibitemShut {NoStop}%
\bibitem [{\citenamefont {Qi}\ \emph {et~al.}(1999)\citenamefont {Qi},
  \citenamefont {Cagin}, \citenamefont {Kimura},\ and\ \citenamefont
  {Goddard}}]{PhysRevB.59.3527}%
  \BibitemOpen
  \bibfield  {author} {\bibinfo {author} {\bibfnamefont {Y.}~\bibnamefont
  {Qi}}, \bibinfo {author} {\bibfnamefont {T.}~\bibnamefont {Cagin}}, \bibinfo
  {author} {\bibfnamefont {Y.}~\bibnamefont {Kimura}}, \ and\ \bibinfo {author}
  {\bibfnamefont {W.~A.}\ \bibnamefont {Goddard}},\ }\href@noop {} {\bibfield
  {journal} {\bibinfo  {journal} {Phys. Rev. B}\ }\textbf {\bibinfo {volume}
  {59}},\ \bibinfo {pages} {3527} (\bibinfo {year} {1999})}\BibitemShut
  {NoStop}%
\bibitem [{\citenamefont {Shibuta}\ and\ \citenamefont
  {Suzuki}(2010{\natexlab{a}})}]{Shibuta:2010eh}%
  \BibitemOpen
  \bibfield  {author} {\bibinfo {author} {\bibfnamefont {Y.}~\bibnamefont
  {Shibuta}}\ and\ \bibinfo {author} {\bibfnamefont {T.}~\bibnamefont
  {Suzuki}},\ }\href@noop {} {\bibfield  {journal} {\bibinfo  {journal} {Chem.\
  Phys.\ Lett.}\ }\textbf {\bibinfo {volume} {498}},\ \bibinfo {pages} {323}
  (\bibinfo {year} {2010}{\natexlab{a}})}\BibitemShut {NoStop}%
\bibitem [{\citenamefont {Shibuta}\ and\ \citenamefont
  {Suzuki}(2011)}]{Shibuta:2011}%
  \BibitemOpen
  \bibfield  {author} {\bibinfo {author} {\bibfnamefont {Y.}~\bibnamefont
  {Shibuta}}\ and\ \bibinfo {author} {\bibfnamefont {T.}~\bibnamefont
  {Suzuki}},\ }\href@noop {} {\bibfield  {journal} {\bibinfo  {journal} {Chem.\
  Phys.\ Lett.}\ }\textbf {\bibinfo {volume} {502}},\ \bibinfo {pages} {82}
  (\bibinfo {year} {2011})}\BibitemShut {NoStop}%
\bibitem [{\citenamefont {III}\ \emph {et~al.}(2007)\citenamefont {III},
  \citenamefont {Brenner}, \citenamefont {Lyshevski},\ and\ \citenamefont
  {(eds.)}}]{Goddard07}%
  \BibitemOpen
  \bibfield  {author} {\bibinfo {author} {\bibfnamefont {W.~G.}\ \bibnamefont
  {III}}, \bibinfo {author} {\bibfnamefont {D.}~\bibnamefont {Brenner}},
  \bibinfo {author} {\bibfnamefont {S.}~\bibnamefont {Lyshevski}}, \ and\
  \bibinfo {author} {\bibfnamefont {G.~I.}\ \bibnamefont {(eds.)}},\
  }\href@noop {} {\emph {\bibinfo {title} {Handbook\ of\ Nanoscience,\
  Engineering,\ and\ Technology}}}\ (\bibinfo  {publisher} {CRC Press},\
  \bibinfo {year} {2007})\BibitemShut {NoStop}%
\bibitem [{\citenamefont {Harutyunyan}\ \emph {et~al.}(2005)\citenamefont
  {Harutyunyan}, \citenamefont {Tokune},\ and\ \citenamefont
  {Mora}}]{Harutyunyan05}%
  \BibitemOpen
  \bibfield  {author} {\bibinfo {author} {\bibfnamefont {A.}~\bibnamefont
  {Harutyunyan}}, \bibinfo {author} {\bibfnamefont {T.}~\bibnamefont {Tokune}},
  \ and\ \bibinfo {author} {\bibfnamefont {E.}~\bibnamefont {Mora}},\
  }\href@noop {} {\bibfield  {journal} {\bibinfo  {journal} {Appl. Phys.
  Lett.}\ }\textbf {\bibinfo {volume} {87}},\ \bibinfo {pages} {051919}
  (\bibinfo {year} {2005})}\BibitemShut {NoStop}%
\bibitem [{\citenamefont {Obolensky}\ \emph {et~al.}(2007)\citenamefont
  {Obolensky}, \citenamefont {Solov'yov}, \citenamefont {Solov'yov},\ and\
  \citenamefont {Greiner}}]{Obolensky07}%
  \BibitemOpen
  \bibfield  {author} {\bibinfo {author} {\bibfnamefont {O.}~\bibnamefont
  {Obolensky}}, \bibinfo {author} {\bibfnamefont {I.}~\bibnamefont
  {Solov'yov}}, \bibinfo {author} {\bibfnamefont {A.}~\bibnamefont
  {Solov'yov}}, \ and\ \bibinfo {author} {\bibfnamefont {W.}~\bibnamefont
  {Greiner}},\ }in\ \href@noop {} {\emph {\bibinfo {booktitle} {International\
  Symposium\ "Atomic\ Cluster\ Collisions:\ structure\ and\ dynamics\ from\
  the\ nuclear\ to\ the\ biological\ scale"}}},\ Vol.\ \bibinfo {volume}
  {31D},\ \bibinfo {editor} {edited by\ \bibinfo {editor} {\bibfnamefont
  {A.}~\bibnamefont {Solov'yov}}}\ (\bibinfo  {publisher} {European Physical
  Society},\ \bibinfo {year} {2007})\ p.\ \bibinfo {pages} {176}\BibitemShut
  {NoStop}%
\bibitem [{\citenamefont {Harutyunyan}\ \emph {et~al.}(2008)\citenamefont
  {Harutyunyan}, \citenamefont {Awasthi}, \citenamefont {Jiang}, \citenamefont
  {Setyawan}, \citenamefont {Mora}, \citenamefont {Tokune}, \citenamefont
  {Bolton},\ and\ \citenamefont {Curtarolo}}]{Harutyunyan08}%
  \BibitemOpen
  \bibfield  {author} {\bibinfo {author} {\bibfnamefont {A.}~\bibnamefont
  {Harutyunyan}}, \bibinfo {author} {\bibfnamefont {N.}~\bibnamefont
  {Awasthi}}, \bibinfo {author} {\bibfnamefont {A.}~\bibnamefont {Jiang}},
  \bibinfo {author} {\bibfnamefont {W.}~\bibnamefont {Setyawan}}, \bibinfo
  {author} {\bibfnamefont {E.}~\bibnamefont {Mora}}, \bibinfo {author}
  {\bibfnamefont {T.}~\bibnamefont {Tokune}}, \bibinfo {author} {\bibfnamefont
  {K.}~\bibnamefont {Bolton}}, \ and\ \bibinfo {author} {\bibfnamefont
  {S.}~\bibnamefont {Curtarolo}},\ }\href@noop {} {\bibfield  {journal}
  {\bibinfo  {journal} {Phys. Rev. Lett.}\ }\textbf {\bibinfo {volume} {100}},\
  \bibinfo {pages} {195502} (\bibinfo {year} {2008})}\BibitemShut {NoStop}%
\bibitem [{\citenamefont {Solov'yov}\ \emph {et~al.}(2012)\citenamefont
  {Solov'yov}, \citenamefont {Yakubovich}, \citenamefont {Nikolaev},
  \citenamefont {Volkovets},\ and\ \citenamefont {Solov'yov}}]{MBNExplorer}%
  \BibitemOpen
  \bibfield  {author} {\bibinfo {author} {\bibfnamefont {I.}~\bibnamefont
  {Solov'yov}}, \bibinfo {author} {\bibfnamefont {A.}~\bibnamefont
  {Yakubovich}}, \bibinfo {author} {\bibfnamefont {P.}~\bibnamefont
  {Nikolaev}}, \bibinfo {author} {\bibfnamefont {I.}~\bibnamefont {Volkovets}},
  \ and\ \bibinfo {author} {\bibfnamefont {A.}~\bibnamefont {Solov'yov}},\
  }\href {\doibase 10.1002/jcc.23086} {\bibfield  {journal} {\bibinfo
  {journal} {J. Comp. Chem., doi 10.1002/jcc.23086}\ } (\bibinfo {year}
  {2012})}\BibitemShut {NoStop}%
\bibitem [{\citenamefont {Nayak}\ \emph {et~al.}(1997)\citenamefont {Nayak},
  \citenamefont {Khanna}, \citenamefont {Rao},\ and\ \citenamefont
  {Jena}}]{Nayak97}%
  \BibitemOpen
  \bibfield  {author} {\bibinfo {author} {\bibfnamefont {S.}~\bibnamefont
  {Nayak}}, \bibinfo {author} {\bibfnamefont {S.}~\bibnamefont {Khanna}},
  \bibinfo {author} {\bibfnamefont {B.}~\bibnamefont {Rao}}, \ and\ \bibinfo
  {author} {\bibfnamefont {P.}~\bibnamefont {Jena}},\ }\href@noop {} {\bibfield
   {journal} {\bibinfo  {journal} {J.~Phys.\ Chem.~A}\ }\textbf {\bibinfo
  {volume} {101}},\ \bibinfo {pages} {1072} (\bibinfo {year}
  {1997})}\BibitemShut {NoStop}%
\bibitem [{\citenamefont {Daw}\ and\ \citenamefont {Baskes}(1983)}]{Daw83}%
  \BibitemOpen
  \bibfield  {author} {\bibinfo {author} {\bibfnamefont {M.}~\bibnamefont
  {Daw}}\ and\ \bibinfo {author} {\bibfnamefont {M.}~\bibnamefont {Baskes}},\
  }\href@noop {} {\bibfield  {journal} {\bibinfo  {journal} {Phys. Rev. Lett.}\
  }\textbf {\bibinfo {volume} {50}},\ \bibinfo {pages} {1285} (\bibinfo {year}
  {1983})}\BibitemShut {NoStop}%
\bibitem [{\citenamefont {Daw}\ and\ \citenamefont {Baskes}(1984)}]{Daw84}%
  \BibitemOpen
  \bibfield  {author} {\bibinfo {author} {\bibfnamefont {M.}~\bibnamefont
  {Daw}}\ and\ \bibinfo {author} {\bibfnamefont {M.}~\bibnamefont {Baskes}},\
  }\href@noop {} {\bibfield  {journal} {\bibinfo  {journal} {Phys. Rev. B}\
  }\textbf {\bibinfo {volume} {29}},\ \bibinfo {pages} {6443} (\bibinfo {year}
  {1984})}\BibitemShut {NoStop}%
\bibitem [{\citenamefont {Finnis}\ and\ \citenamefont
  {Sinclair}(1984)}]{Finnis84}%
  \BibitemOpen
  \bibfield  {author} {\bibinfo {author} {\bibfnamefont {M.}~\bibnamefont
  {Finnis}}\ and\ \bibinfo {author} {\bibfnamefont {J.}~\bibnamefont
  {Sinclair}},\ }\href@noop {} {\bibfield  {journal} {\bibinfo  {journal}
  {Philos. Mag.}\ }\textbf {\bibinfo {volume} {50}},\ \bibinfo {pages} {45}
  (\bibinfo {year} {1984})}\BibitemShut {NoStop}%
\bibitem [{\citenamefont {Foiles}\ \emph {et~al.}(1986)\citenamefont {Foiles},
  \citenamefont {Daw},\ and\ \citenamefont {Baskes}}]{Foiles86}%
  \BibitemOpen
  \bibfield  {author} {\bibinfo {author} {\bibfnamefont {S.}~\bibnamefont
  {Foiles}}, \bibinfo {author} {\bibfnamefont {M.}~\bibnamefont {Daw}}, \ and\
  \bibinfo {author} {\bibfnamefont {M.}~\bibnamefont {Baskes}},\ }\href@noop {}
  {\bibfield  {journal} {\bibinfo  {journal} {Phys. Rev. B}\ }\textbf {\bibinfo
  {volume} {33}},\ \bibinfo {pages} {7983} (\bibinfo {year}
  {1986})}\BibitemShut {NoStop}%
\bibitem [{\citenamefont {Sutton}\ and\ \citenamefont {Chen}(1990)}]{Sutton90}%
  \BibitemOpen
  \bibfield  {author} {\bibinfo {author} {\bibfnamefont {A.}~\bibnamefont
  {Sutton}}\ and\ \bibinfo {author} {\bibfnamefont {J.}~\bibnamefont {Chen}},\
  }\href@noop {} {\bibfield  {journal} {\bibinfo  {journal} {Philos.\ Mag.\
  Lett.}\ }\textbf {\bibinfo {volume} {61}},\ \bibinfo {pages} {139} (\bibinfo
  {year} {1990})}\BibitemShut {NoStop}%
\bibitem [{\citenamefont {Sutton}\ \emph {et~al.}(1996)\citenamefont {Sutton},
  \citenamefont {Godwin},\ and\ \citenamefont {Horsfield}}]{Sutton96}%
  \BibitemOpen
  \bibfield  {author} {\bibinfo {author} {\bibfnamefont {A.}~\bibnamefont
  {Sutton}}, \bibinfo {author} {\bibfnamefont {P.}~\bibnamefont {Godwin}}, \
  and\ \bibinfo {author} {\bibfnamefont {A.}~\bibnamefont {Horsfield}},\
  }\href@noop {} {\bibfield  {journal} {\bibinfo  {journal} {MRS Bull.}\
  }\textbf {\bibinfo {volume} {21}},\ \bibinfo {pages} {42} (\bibinfo {year}
  {1996})}\BibitemShut {NoStop}%
\bibitem [{\citenamefont {Raffi-Tabar}\ and\ \citenamefont
  {Sutton}(1991)}]{Sutton91}%
  \BibitemOpen
  \bibfield  {author} {\bibinfo {author} {\bibfnamefont {H.}~\bibnamefont
  {Raffi-Tabar}}\ and\ \bibinfo {author} {\bibfnamefont {A.}~\bibnamefont
  {Sutton}},\ }\href@noop {} {\bibfield  {journal} {\bibinfo  {journal}
  {Philos.\ Mag.\ Lett.}\ }\textbf {\bibinfo {volume} {63}},\ \bibinfo {pages}
  {217} (\bibinfo {year} {1991})}\BibitemShut {NoStop}%
\bibitem [{\citenamefont {Todd}\ and\ \citenamefont
  {Lynden-Bell}(1993)}]{Todd93}%
  \BibitemOpen
  \bibfield  {author} {\bibinfo {author} {\bibfnamefont {B.}~\bibnamefont
  {Todd}}\ and\ \bibinfo {author} {\bibfnamefont {R.}~\bibnamefont
  {Lynden-Bell}},\ }\href@noop {} {\bibfield  {journal} {\bibinfo  {journal}
  {Surf.\ Sci.}\ }\textbf {\bibinfo {volume} {287}},\ \bibinfo {pages} {191}
  (\bibinfo {year} {1993})}\BibitemShut {NoStop}%
\bibitem [{\citenamefont {Lynden-Bell}(1995)}]{Lynden95}%
  \BibitemOpen
  \bibfield  {author} {\bibinfo {author} {\bibfnamefont {R.}~\bibnamefont
  {Lynden-Bell}},\ }\href@noop {} {\bibfield  {journal} {\bibinfo  {journal}
  {J.~Phys.:\ Condens.\ Matter}\ }\textbf {\bibinfo {volume} {7}},\ \bibinfo
  {pages} {4603} (\bibinfo {year} {1995})}\BibitemShut {NoStop}%
\bibitem [{\citenamefont {Doye}\ and\ \citenamefont {Wales}(1998)}]{Doye98}%
  \BibitemOpen
  \bibfield  {author} {\bibinfo {author} {\bibfnamefont {J.}~\bibnamefont
  {Doye}}\ and\ \bibinfo {author} {\bibfnamefont {D.}~\bibnamefont {Wales}},\
  }\href@noop {} {\bibfield  {journal} {\bibinfo  {journal} {New J. Chem.}\
  }\textbf {\bibinfo {volume} {22}},\ \bibinfo {pages} {733} (\bibinfo {year}
  {1998})}\BibitemShut {NoStop}%
\bibitem [{\citenamefont {Landau}\ and\ \citenamefont {Lifshitz}(1980)}]{LL5}%
  \BibitemOpen
  \bibfield  {author} {\bibinfo {author} {\bibfnamefont {L.}~\bibnamefont
  {Landau}}\ and\ \bibinfo {author} {\bibfnamefont {E.}~\bibnamefont
  {Lifshitz}},\ }\href@noop {} {\emph {\bibinfo {title} {Statistical physics,
  Part I}}}\ (\bibinfo  {publisher} {Butterworth-Heinemann, Oxford},\ \bibinfo
  {year} {1980})\BibitemShut {NoStop}%
\bibitem [{\citenamefont {Kirkpatrick}\ \emph {et~al.}(1983)\citenamefont
  {Kirkpatrick}, \citenamefont {Gelatt},\ and\ \citenamefont
  {Vecchi}}]{Kirkpatrick:1983zz}%
  \BibitemOpen
  \bibfield  {author} {\bibinfo {author} {\bibfnamefont {S.}~\bibnamefont
  {Kirkpatrick}}, \bibinfo {author} {\bibfnamefont {C.}~\bibnamefont {Gelatt}},
  \ and\ \bibinfo {author} {\bibfnamefont {M.}~\bibnamefont {Vecchi}},\
  }\href@noop {} {\bibfield  {journal} {\bibinfo  {journal} {Science}\ }\textbf
  {\bibinfo {volume} {220}},\ \bibinfo {pages} {671} (\bibinfo {year}
  {1983})}\BibitemShut {NoStop}%
\bibitem [{\citenamefont {Dick}\ \emph {et~al.}(2011)\citenamefont {Dick},
  \citenamefont {Solov'yov},\ and\ \citenamefont {Solov'yov}}]{Dick11}%
  \BibitemOpen
  \bibfield  {author} {\bibinfo {author} {\bibfnamefont {V.}~\bibnamefont
  {Dick}}, \bibinfo {author} {\bibfnamefont {I.}~\bibnamefont {Solov'yov}}, \
  and\ \bibinfo {author} {\bibfnamefont {A.}~\bibnamefont {Solov'yov}},\
  }\href@noop {} {\bibfield  {journal} {\bibinfo  {journal} {Phys. Rev. B}\
  }\textbf {\bibinfo {volume} {84}},\ \bibinfo {pages} {115408} (\bibinfo
  {year} {2011})}\BibitemShut {NoStop}%
\bibitem [{\citenamefont {Shibuta}\ and\ \citenamefont
  {Suzuki}(2010{\natexlab{b}})}]{Shibuta10}%
  \BibitemOpen
  \bibfield  {author} {\bibinfo {author} {\bibfnamefont {Y.}~\bibnamefont
  {Shibuta}}\ and\ \bibinfo {author} {\bibfnamefont {T.}~\bibnamefont
  {Suzuki}},\ }\href@noop {} {\bibfield  {journal} {\bibinfo  {journal} {Chem.\
  Phys.\ Lett.}\ }\textbf {\bibinfo {volume} {498}},\ \bibinfo {pages} {323}
  (\bibinfo {year} {2010}{\natexlab{b}})}\BibitemShut {NoStop}%
\end{thebibliography}
\end{document}